\documentclass[showpacs,floatfix,amssymb]{article}
\usepackage{graphicx,subfigure,amsmath,color,amsmath,amssymb,hyperref,cite}
\begin{document}
\title{The Texture One Zero Neutrino Mass Matrix  
With Vanishing Trace}

\author{Madan Singh$^{*} $\\
\it Department of Physics, National Institute of Technology Kurukshetra,\\
\it Haryana,136119, India.\\
\it $^{*}$singhmadan179@gmail.com
}
\maketitle
\begin{abstract}
In the light of latest neutrino oscillation data, we have investigated the one zero Majorana neutrino mass matrix $M_{\nu}$ with zero sum condition of mass eigen values in the flavor basis, where charged lepton mass matrix is diagonal.  Among the six possible one-zero cases,  it is found that only five can survive the current experimental data,  while case with (1, 1) vanishing element of $M_{\nu}$ is ruled out, if zero trace condition is imposed at 3$\sigma$ confidence level (CL). Numerical and some approximate analytical results are presented.
\end{abstract}

\section{Introduction}
The Double Chooz, Daya Bay and RENO Collaborations \cite{1,2,3}, have finally established the non-zero and relatively large value of the reactor mixing angle $\theta_{13}$, hence the number of known available  neutrino oscillation parameters approaches to five viz. two mass squared differences ($\delta m^{2}$, $\Delta m^{2}$) and three neutrino mixing angles ($\theta_{12}$, $\theta_{23}$, $\theta_{13}$). However, any general 3 $\times$ 3 neutrino mass matrix contains more parameters than can be measured in realistic experiments. In fact, assuming the Majorana type nature of neutrinos, the neutrino mass matrix contains nine real free parameters: three neutrino masses ($m_{1}, m_{2}, m_{3}$), three flavor mixing angles ($\theta_{12}$, $\theta_{23}$, $\theta_{13}$) and three CP violating phases ($\delta, \rho, \sigma$). \\
In order to reduce the number of free parameters, several phenomenological ideas, in particular texture zeros \cite{4,5,6,7,8,9,10,11,12} have been widely adopted in the literature. The imposition of texture zeros in neutrino mass matrix leads to some important phenomenological relations between flavor mixing angles and fermion mass ratios \cite{10,11,12}. In the flavor basis, where charged lepton mass matrix is diagonal, at the most two zeros are allowed in neutrino mass matrix, which are consistent with neutrino oscillation data \cite{12}. 
The analysis of two texture zero neutrino mass matrices limits the number of experimentally viable cases to seven. The phenomenological implications of one texture zero neutrino mass matrices have also been studied in the literature \cite{7,8,9} and it has been observed that all the six  cases are viable with experimental data. However, the imposition of  single texture zero condition in neutrino mass matrix makes available larger parametric space for viability with the data compared with texture two zero.  In order to impart predictability to texture one zero, additional constraints in the form of vanishing  determinant \cite{13} or trace  can be incorporated. The phenomenological implication of determinantless condition on texture one zero have been rigoursly studied in Refs. \cite{7,13,14}.  The implication of traceless condition, was first put forward in \cite{15} wherein the anomalies of solar and atmospheric neutrino oscillation experiments as well as the LSND experiment were simulataneously explained in the framework of three neutrinos. In Ref. \cite{16}, X. G. He and  A. Zee  have particulary investigated the case of CP conserving traceless neutrino mass matrix for explaining the solar  and atmospheric neutrino deficits. Further motivation of traceless mass matrices can be provided by models wherein neutrino mass matrix can be constructed through a commutator of two matrices, as it happens in models of radiative mass generation \cite{17}. In Ref. \cite{18},  H. Alhendi et. al. have studied the case of  two tracless  submatrices of Majorana mass matrix in the flavor basis and carried out a detailed numerical analysis at 3$\sigma$ confidence level. The phenomenological implications of traceless neutrino mass matrix on neutrino masses, CP violating phases and effective neutrino mass term is also studied in Ref. \cite{19}, for both normal and inverted mass ordering and in case of CP conservation and violation, respectively. In the present work we impose the traceless condition on texture one zero Majorana mass matrix and investigate  the outcomes of such condition on the parameteric space of neutrino masses $(m_{1}, m_{2}, m_{3}$) and  CP violating phases ($\delta, \rho, \sigma$).   

Assuming the Majorana nature of neutrinos, neutrino mass matrix is complex symmetric. In the flavor basis, if one of the element is considered to be zero, the number of possible cases turns out to be six, which are given below 

\begin{equation}\label{eq1}
T_{1}:\left(
\begin{array}{ccc}
0 & \times & \times\\
 \times & \times & \times\\
\times& \times & \times \\
\end{array}
\right), T_{2}:\left(
\begin{array}{ccc}
\times& \times  & \times  \\
\times & 0 & \times\\
\times& \times & \times \\
\end{array}
\right), T_{3}:\left(
\begin{array}{ccc}
    \times& \times & \times \\
   \times & \times & \times\\
  \times&\times & 0 \\
\end{array}
\right);
\end{equation}
\begin{equation}\label{eq2}
\medskip
  T_{4}:\left(
\begin{array}{ccc}
    \times& 0 & \times  \\
  0 & \times & \times\\
  \times& \times & \times \\
\end{array}
\right), T_{5}:\left(
\begin{array}{ccc}
    \times& \times & 0 \\
  \times & \times & \times \\
  0& \times & \times \\
\end{array}
\right),
T_{6}: \left(
\begin{array}{ccc}
   \times& \times& \times  \\
   \times& \times & 0\\
  \times& 0 & \times \\
\end{array}
\right);
\end{equation}
\begin{flushleft}
where '$\times$ ' stands for non-zero element and complex matrix
element.
\end{flushleft}
Among these possible cases, there exists a permutation symmetry between certain pair of cases viz. ($T_{2}$, $T_{3}$) and ($T_{4}$, $T_{5}$), while cases $T_{1}$ and $T_{6}$ tranform onto themselves independently. The origin of permutation symmetry is explained from the fact that these pairs are related by exchange of 2-3 rows and 2-3 columns of neutrino mass matrix. The corresponding permutation matrix is given by\\

\qquad\qquad \qquad \qquad \qquad \qquad $P_{23}=\left(
\begin{array}{ccc}
    1& 0& 0 \\
  0 & 0 & 1\\
  0& 1& 0 \\
\end{array}
\right)$,\\
which leads to the following relations among the neutrino oscillation parameters 
\begin{equation}\label{eq3}
\theta_{12}^{X}=\theta_{12}^{Y},\;
\theta_{23}^{X}=90^{0}-\theta_{23}^{Y},\;
\theta_{13}^{X}=\theta_{13}^{Y}, \delta^{X}=\delta^{Y}\pm 180^{0},
\end{equation}
where X and Y superscripts denote the cases related by 2-3 permutation symmetry. 

The rest of the work is planned as follows: In Section 2, we discuss the methodology used to reconstruct the Majorana neutrino mass matrix and subsequently obtain some useful phenomenological relations of neutrino mass ratios and Majorana phases by incorporating texture one zero and zero trace conditions simultaneously. In Section 3, we present the numerical analysis using some approximate analytical relations.  In Section 4, we summarize our work.

\section{Formalism}
 In the flavor basis, the Majorana neutrino mass matrix $M_{\nu}$, depending on three neutrino masses $(m_{1}, m_{2}, m_{3})$ and the flavor mixing matrix  can be expressed as,
\begin{equation}\label{eq4}
 M_{\nu}=V\left(
\begin{array}{ccc}
    m_{1}& 0& 0 \\
  0 & m_{2} & 0\\
  0& 0& m_{3} \\
  \end{array}
  \right)V^{T}.
\end{equation}
The mixing matrix V can be written as $V=UP$,
where $U$ denotes the neutrino mixing matrix
consisting of three flavor mixing angles and one
Dirac-like CP violating phase, whereas, the
matrix $P$ is a diagonal phase matrix, i.e.,
$P$=diag($e^{i\rho},e^{i\sigma},1$) with $\rho$
and $\sigma$ being the two Majorana CP violating
phases. The neutrino mass matrix $M_{\nu}$ can
then be rewritten as
\begin{equation}\label{eq5}
M_{\nu}=\left(
\begin{array}{ccc}
   M_{ee}& M_{e \mu}& M_{e \tau} \\
  M_{e \mu} & M_{\mu \mu} & M_{\mu \tau}\\
  M_{e \tau}& M_{\mu \tau}& M_{\tau \tau} \\
  \end{array}
  \right)=U\left(
\begin{array}{ccc}
    \lambda_{1}& 0& 0 \\
  0 & \lambda_{2} & 0\\
  0& 0& \lambda_{3} \\
\end{array}
\right) U^{T},
\end{equation}\\
where $\lambda_{1} = m_{1} e^{2i\rho},\lambda_{2}
= m_{2} e^{2i\sigma} ,\lambda_{3} = m_{3}.$

For the purpose of calculations, we have adopted
the parameterization of the mixing matrix $U$
considered by Ref. \cite{5}, e.g.,
\begin{equation}\label{eq6}
U=\left(
\begin{array}{ccc}
 c_{12}c_{13}& s_{12}c_{13}& s_{13} \\
-c_{12}s_{23}s_{13}-s_{12}c_{23}e^{-i\delta} &
-s_{12}s_{23}s_{13}+c_{12}c_{23}e^{-i\delta} &
s_{23}c_{13}\\
 -c_{12}c_{23}s_{13}+s_{12}s_{23}e^{-i\delta}
 & -s_{12}c_{23}s_{13}-c_{12}s_{23}e^{-i\delta}& c_{23}c_{13} \\
\end{array}
 \right),
\end{equation}\\
where $c_{ij} = cos \theta_{ij}, s_{ij}= sin
\theta_{ij}$ for i,j=1,2,3 and $\delta$ is the CP
violating phase.\\
If one of the elements of $M_{\nu}$ is considered zero, i.e. $M_{lm} = 0$, it leads to the following constraint equation
\begin{equation}\label{eq7}
U_{l1}U_{m1}\lambda_{1}+U_{l2}U_{m2}\lambda_{2}+
U_{l3}U_{m3} \lambda_{3}=0,
\end{equation}
where l, m run over e, $\mu$ and $\tau$. \\
The traceless condition implies sum of the mass eigen values in neutrino mass matrix is zero i.e.
\begin{equation}\label{eq8}
\lambda_{1}+\lambda_{2}+\lambda_{3}=0.
\end{equation}
Using Eqs. \ref{eq7} and \ref{eq8}, we obtain
\begin{equation}\label{eq9}
\dfrac{\lambda_{1}}{\lambda_{3}}=\dfrac{U_{l2}U_{m2}-U_{l3}U_{m3}}{U_{l1}U_{m1}-U_{l2}U_{m2}},
\end{equation}
\begin{equation}\label{eq10}
\dfrac{\lambda_{2}}{\lambda_{3}}=\dfrac{U_{l3}U_{m3}-U_{l1}U_{m1}}{U_{l1}U_{m1}-U_{l2}U_{m2}}.
\end{equation}\\
The magnitudes of neutrino mass ratios are given by
\begin{equation}\label{eq11}
\xi \equiv \dfrac{m_{1}}{m_{3}} =\bigg|\dfrac{U_{l2}U_{m2}-U_{l3}U_{m3}}{U_{l1}U_{m1}-U_{l2}U_{m2}}\bigg|, 
\end{equation} 
\begin{equation}\label{eq12}
\zeta \equiv \dfrac{m_{2}}{m_{3}} =\bigg|\dfrac{U_{l3}U_{m3}-U_{l1}U_{m1}}{U_{l1}U_{m1}-U_{l2}U_{m2}}\bigg|. 
\end{equation}\\
Using Eqs. \ref{eq9} and \ref{eq10}, we find the following analytical relations for Majorana phases  ($\rho, \sigma$) 
\begin{equation}\label{eq13}
\rho=\dfrac{1}{2}arg \left(\dfrac{U_{l2}U_{m2}-U_{l3}U_{m3}}{U_{l1}U_{m1}-U_{l2}U_{m2}}\right),  
\end{equation}

\begin{equation}\label{eq14}
\sigma=\dfrac{1}{2}arg \left(\dfrac{U_{l3}U_{m3}-U_{l1}U_{m1}}{U_{l1}U_{m1}-U_{l2}U_{m2}} \right).
\end{equation}
Thus neutrino mass ratios ($\xi, \zeta$) and two Majorana-type CP-violating phases ($\rho, \sigma$) can fully be determined in terms of three mixing angles ($\theta_{12}$, $\theta_{23}$, $\theta_{13}$)and the Dirac-type CP violating phase ($\delta$). The ratio of two neutrino mass-squared differences in terms of neutrino mass ratios $\xi$ and $\zeta$ is given by
\begin{equation}\label{eq15}
 R_{\nu}=\frac{\delta m^{2}}{|\Delta m^{2}|}=\frac{2(\zeta ^{2}-\xi ^{2})}{\left |2-(\zeta ^{2}+\xi ^{2})  \right |},
 \end{equation}\\
 where $\delta m^{2}=(m_{2}^{2}-m_{1}^{2})$ and $\Delta m^{2}=|m_{3}^{2}-\frac{1}{2}(m_{1}^{2}+m_{2}^{2})|$ \cite{20} corresponds to solar  and atmospheric neutrino squared differences, respectively. The sign of $\Delta m^{2}$ is still not known experimentally i.e. $\Delta m^{2}>0$ or $\Delta m^{2}<0$ corresponds to the normal or inverted mass ordering  of neutrinos. \\
The expressions for three neutrino masses ($m_{1}, m_{2}, m_{3}$) can be given  as 
 \begin{equation}\label{eq16}
 m_{3}=\sqrt{\frac{\delta m^{2}}{(\zeta ^{2}-\xi ^{2})}}, \qquad m_{2}=m_{3}\zeta ,  \qquad m_{1}=m_{3} \xi.
 \end{equation}
  Thus the neutrino mass spectrum can be fully determined.\\
 The expression for Jarlskog rephasing parameter $J_{CP}$, which is a measure of CP violation, is given by
\begin{equation}\label{eq17}
J_{CP}=s_{12} c_{12} s_{23}  c_{23} s_{13}  c_{13}^{2} \sin\delta.
\end{equation}
\begin{table}
\begin{small}
\begin{center}
\begin{tabular}{|c|c|c|c|c|}
  \hline
  Parameter& Best Fit & 1$\sigma$ & 2$\sigma$ & 3$\sigma$ \\
  \hline
   $\delta m^{2}$ $[10^{-5}eV^{2}]$ & $7.60$& $7.42$ - $7.79$ & $7.26$ - $7.99$ & $7.11$ - $8.18$ \\
   \hline
   $|\Delta m^{2}_{31}|$ $[10^{-3}eV^{2}]$ (NO) & $2.48$ & $2.41$ - $2.53$ & $2.35$ - $2.59$ & $2.30$ - $2.65$\\
   \hline
  $|\Delta m^{2}_{31}|$ $[10^{-3}eV^{2}]$ (IO) & $2.38$ & $2.32$ - $2.43$ & $2.26$ - $2.48$ & $2.20$ - $2.54$ \\
  \hline
  $\theta_{12}$ & $34.6^{\circ}$ & $33.6^{\circ}$ - $35.6^{\circ}$ & $32.7^{\circ}$ - $36.7^{\circ}$ & $31.8^{\circ}$ - $37.8^{\circ}$\\
  \hline
  $ \theta_{23}$ (NO) & $48.9^{\circ}$ &$41.7^{\circ}$ - $50.7^{\circ}$  & $40.0^{\circ}$ - $52.1^{\circ}$ & $38.8^{\circ}$ - $53.3^{\circ}$ \\
  \hline
  $\theta_{23}$ (IO)& $49.2^{\circ}$ & $46.9^{\circ}$ - $50.7^{\circ}$ & $41.3^{\circ}$ - $52.0^{\circ} $& $39.4^{\circ}$ - $53.1^{\circ}$ \\
  \hline
  $\theta_{13}$ (NO) & $8.6^{\circ}$ & $8.4^{\circ}$ - $8.9^{\circ}$ & $8.2^{\circ}$ - $9.1^{\circ}$& $7.9^{\circ}$ - $9.3^{\circ}$ \\
  \hline
  $\theta_{13}$ (IO) & $8.7^{\circ}$ & $8.5^{\circ}$ - $8.9^{\circ}$ & $8.2^{\circ}$ - $9.1^{\circ}$ & $8.0^{\circ}$ - $9.4^{\circ}$ \\
  \hline
  $\delta$ (NO) & $254^{\circ}$ & $182^{\circ}$ - $353^{\circ}$& $0^{\circ}$ - $360^{\circ}$ & $0^{\circ}$ - $360^{\circ}$ \\
  \hline
  $\delta$ (IO) &$266^{\circ}$& $210^{\circ}$ - $322^{\circ}$ & $0^{\circ}$ - $16^{\circ}$ $\oplus$ $ 155^{\circ}$ - $360^{\circ}$  & $0^{\circ}$ - $360^{\circ}$ \\
\hline
\end{tabular}
\caption{\label{tab1} Current neutrino oscillation parameters from global fits at 1$\sigma$, 2$\sigma$ and 3$\sigma$ confidence level \cite{21}. NO(IO) refers to normal (inverted) neutrino mass ordering.}
\end{center}
\end{small}
\end{table}

\section{Numerical results and Discussion}
The experimental constraints on neutrino parameters at 1$\sigma$, 2$\sigma$ and 3$\sigma$ confidence level (CL) are given in Table \ref{tab1}.\\
The effective Majorana mass term relevant for neutrinoless double beta ($0\nu\beta\beta$) decay is given by
\begin{equation}\label{eq18}
|M|_{ee}=|m_{1}c_{12}^{2}c_{13}^{2}e^{2i\rho}+m_{2}s_{12}^{2}c_{13}^{2}e^{2i\sigma}+m_{3}s_{13}^{2}|.
\end{equation}
The future observation of $0\nu\beta\beta$ decay would imply lepton number violation and Majorana character of neutrinos.  For recent reviews see Refs. \cite{22, 23}. There are large number of projects such as CUORICINO \cite{24}, CUORE \cite{25}, GERDA \cite{26}, MAJORANA \cite{27}, SuperNEMO \cite{28}, EXO \cite{29}, GENIUS \cite{30} which target to achieve a sensitivity upto 0.01eV for  $|M|_{ee}$. For the present analysis, we assume the upper limit on $|M|_{ee}$  to be less than 0.5eV at 3$\sigma$ CL \cite{23}. The data collected from the Planck satellite \cite{31} combined with other cosmological data put a limit on the sum of neutrino masses as
\begin{equation}\label{eq19}
\Sigma = \sum_{i = 1}^3 m_{i} < 0.23 \textrm{eV \ \ \ at 95\% CL.}
\end{equation}
 Here, we take rather more conservative limit on sum of neutrino masses $(\Sigma$) (i.e.  $\Sigma <1$eV) at 3$\sigma$ CL.  We span the parameter space of input neutrino oscillation parameters ($\theta_{12}$, $\theta_{23}$, $\theta_{13},\delta m^{2}$, $\Delta m^{2}$) by choosing the randomly generated points of the order of $10^{6-7}$. Using Eq. \ref{eq15}, the parameter space of CP-violating phases ($\delta, \rho, \sigma$), effective mass term  $|M|_{ee}$, neutrino masses ($m_{1}, m_{2}, m_{3}$) can be subsequently constrained. In order to interpret the phenomenological results, some approximate analytical relations (up to certain leading order term of $s_{13}$) have been used in the following discussion. The exact analytical relations of neutrino mass ratios ($\xi, \zeta $) have been provided in Table \ref{tab2}. 
 
 \subsection{Case $T_{1}$}
Using Eqs. \ref{eq6}, \ref{eq9} and \ref{eq10}, in the leading order term  of $\theta_{13}$, we obtain the following analytical relations
\begin{equation}\label{eq20}
 \frac{\lambda_{1}}{\lambda_{3}}\approx \; sec2 \theta_{12} s_{12}^{2},
\end{equation}
\begin{equation}\label{eq21}
 \frac{\lambda_{2}}{\lambda_{3}}\approx -\; sec2 \theta_{12} c_{12}^{2}.
\end{equation}
For NO, using Eq. \ref{eq15}, we obtain  $R_{\nu}\approx \zeta^{2}-\xi^{2} \approx sec2\theta_{12}$.  Using 3$\sigma$ experimental range of oscillation parameters, we find,  $2.23\leq R_{\nu} \leq 4.02$, which excludes the experimental range of  $R_{\nu}$  and for IO , we have
\begin{equation}\label{eq22}
R_{\nu}\approx \frac{sec2 \theta_{12}}{sec^{2}2 \theta_{12} c^{4}_{12}-1},
\end{equation}
which is again inconsistent with current experimental data as $R_{\nu}> 0.75$. Therefore, case $T_{1}$  is ruled out with the latest  neutrino oscillation data at 3$\sigma$ CL.\\

\subsection{Case $T_{2}$}
\begin{figure}[h!]
\begin{center}
\subfigure[]{\includegraphics[width=0.35\columnwidth]{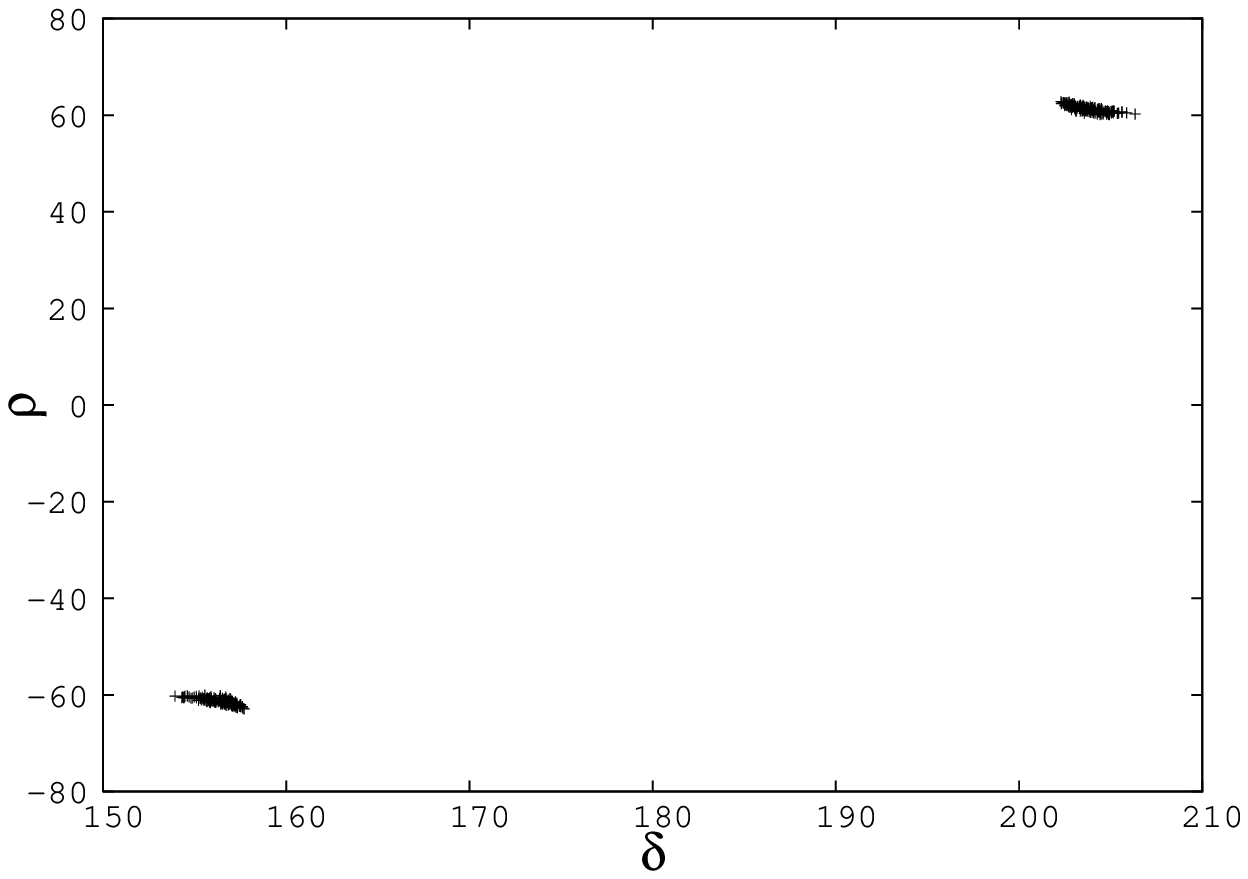}} \ \ \
\subfigure[]{\includegraphics[width=0.35\columnwidth]{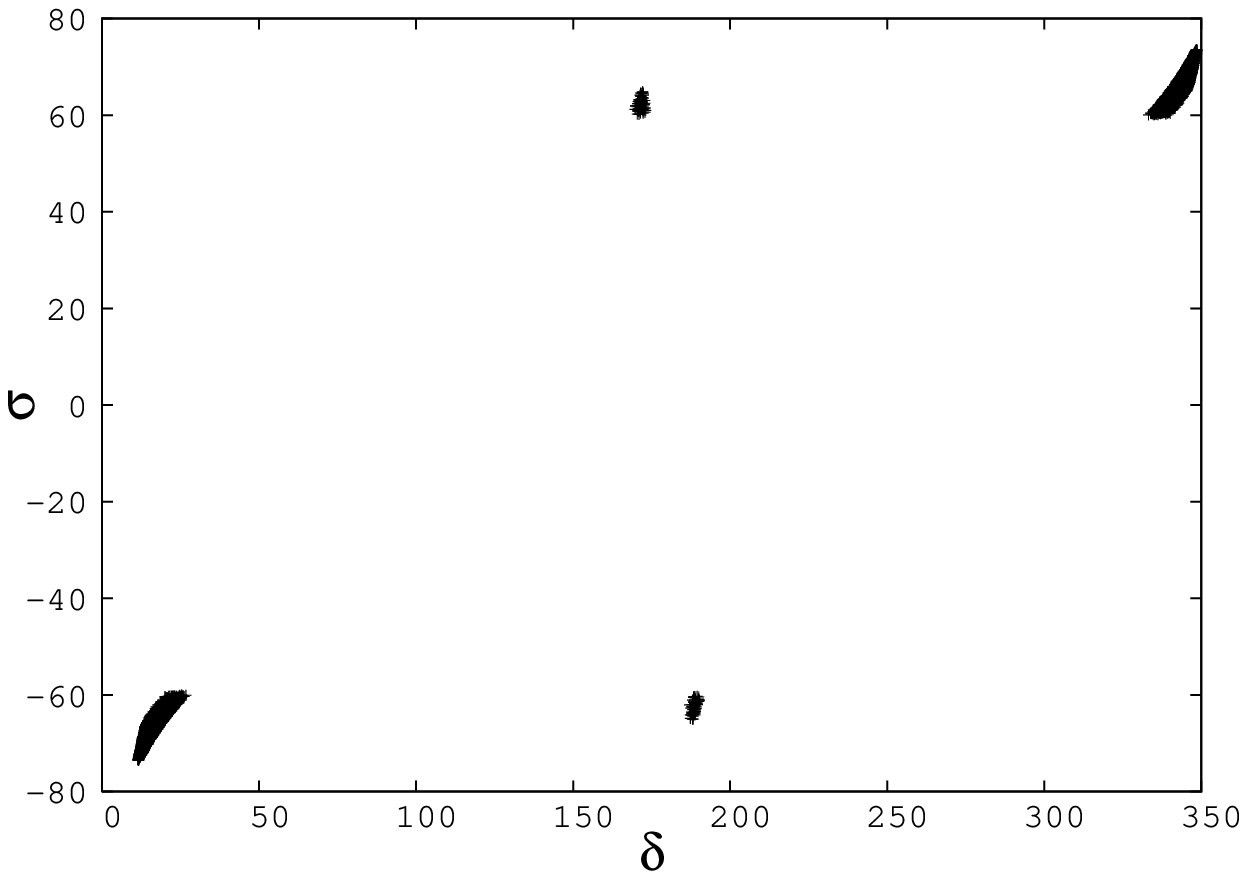}}\\
\subfigure[]{\includegraphics[width=0.35\columnwidth]{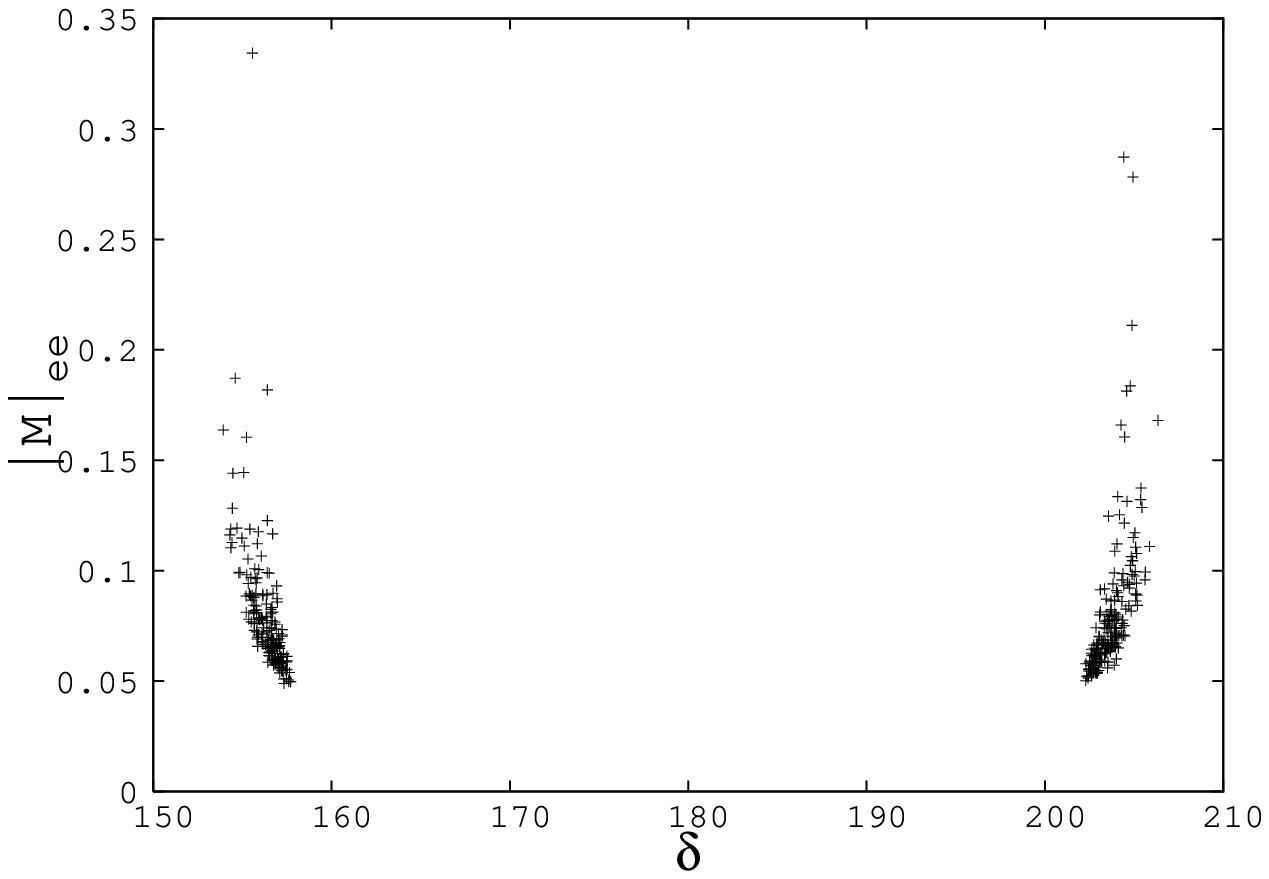}} \ \ \
\subfigure[]{\includegraphics[width=0.35\columnwidth]{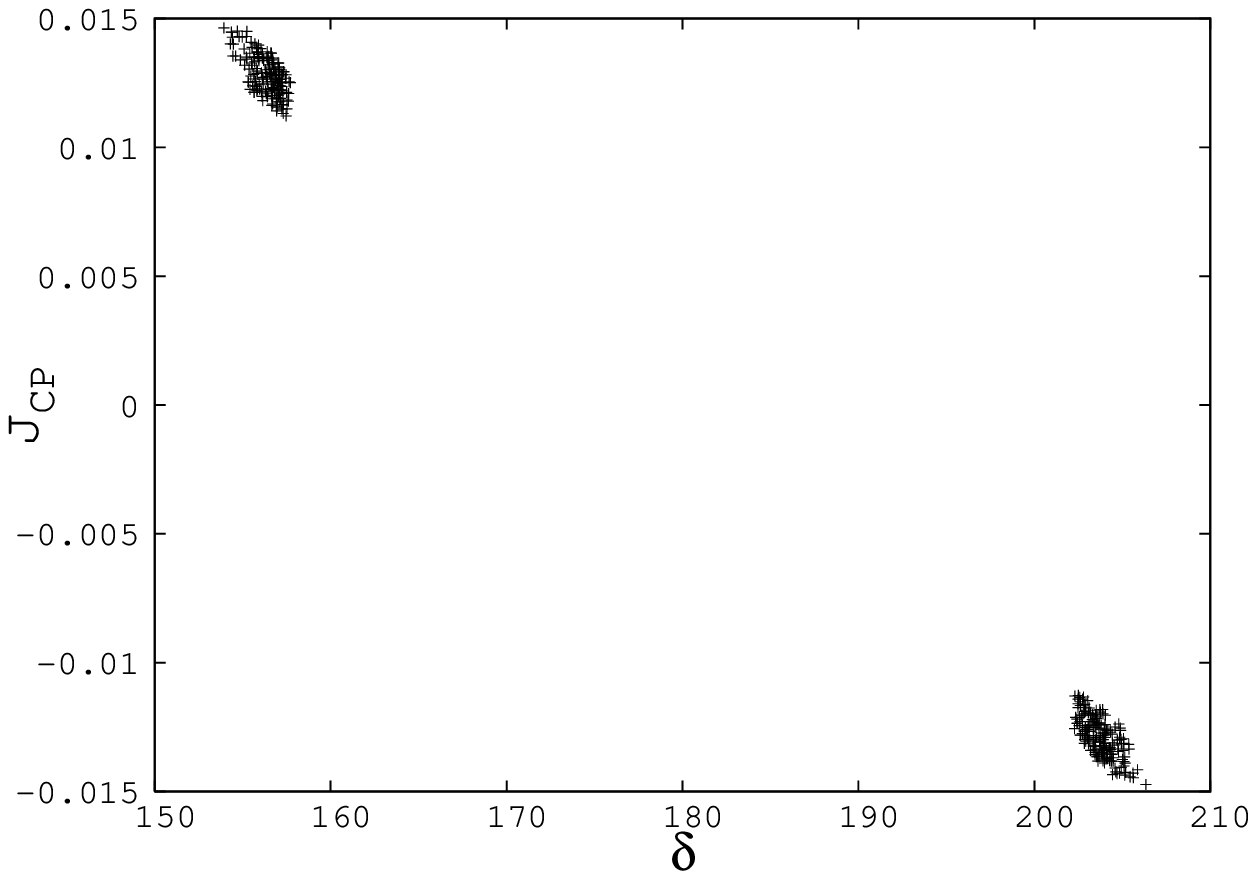}}
\caption{\label{fig1} Case $T_{2}$ (NO):  Scattering plots of  Majorana phases, Dirac CP-violating phase ($\delta$), effective neutrino mass  $|M|_{ee}$,  Jarlskog rephrasing invariant ($J_{CP}$) have been shown. All the phase angles ($\delta, \rho, \sigma$) are measured in degrees and $|M|_{ee}$ is in eV unit.  }
\end{center}
\end{figure}

\begin{figure}[h!]
\begin{center}
\subfigure[]{\includegraphics[width=0.35\columnwidth]{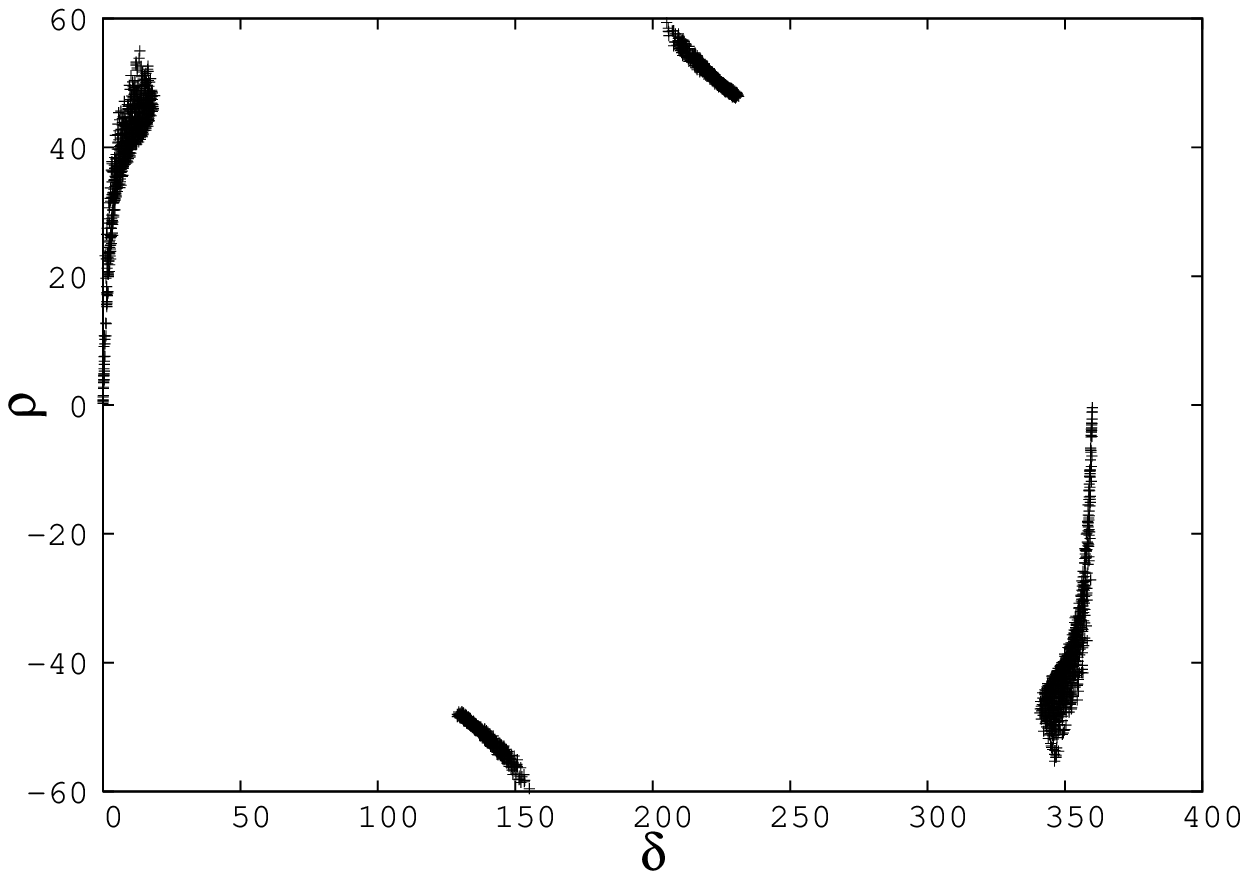}} \ \ \
\subfigure[]{\includegraphics[width=0.35\columnwidth]{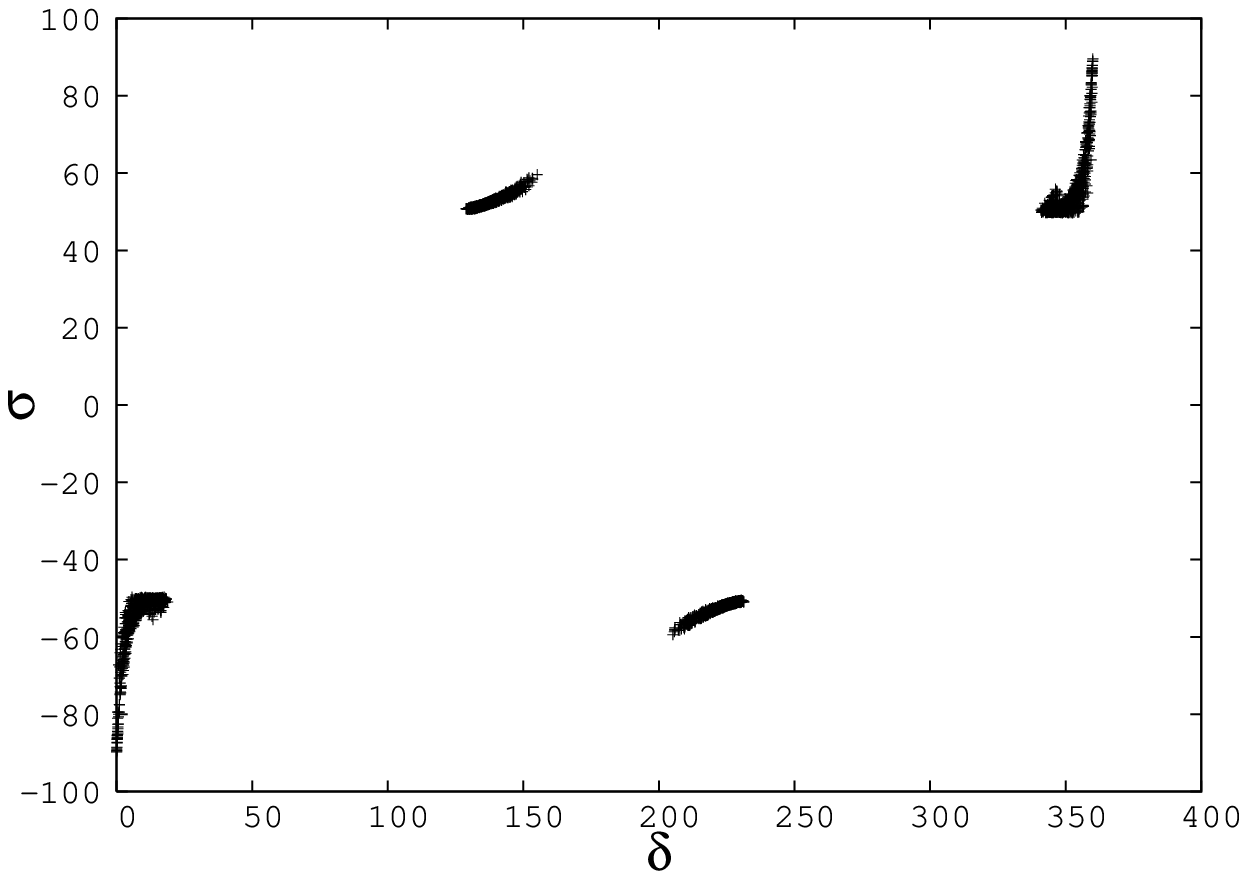}}\\
\subfigure[]{\includegraphics[width=0.35\columnwidth]{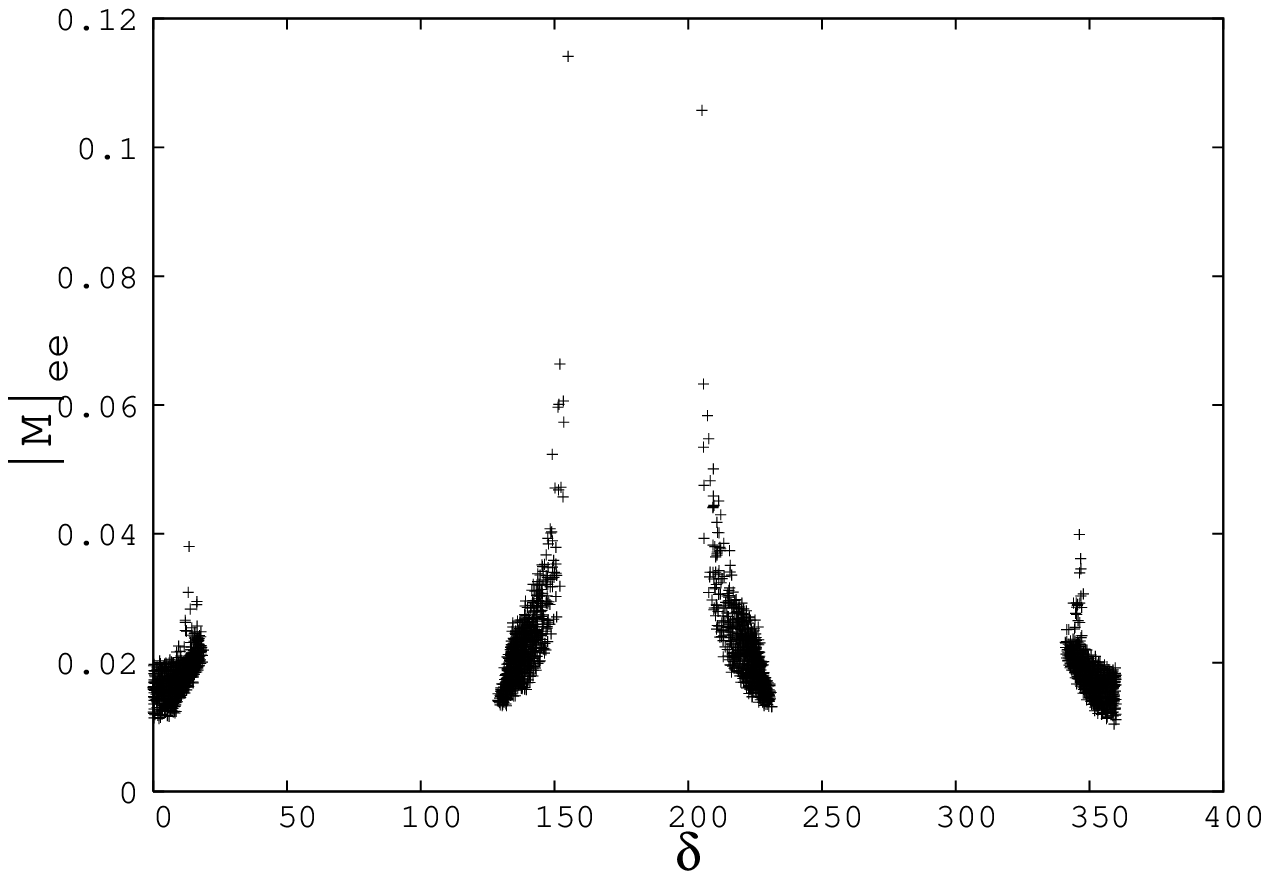}} \ \ \
\subfigure[]{\includegraphics[width=0.35\columnwidth]{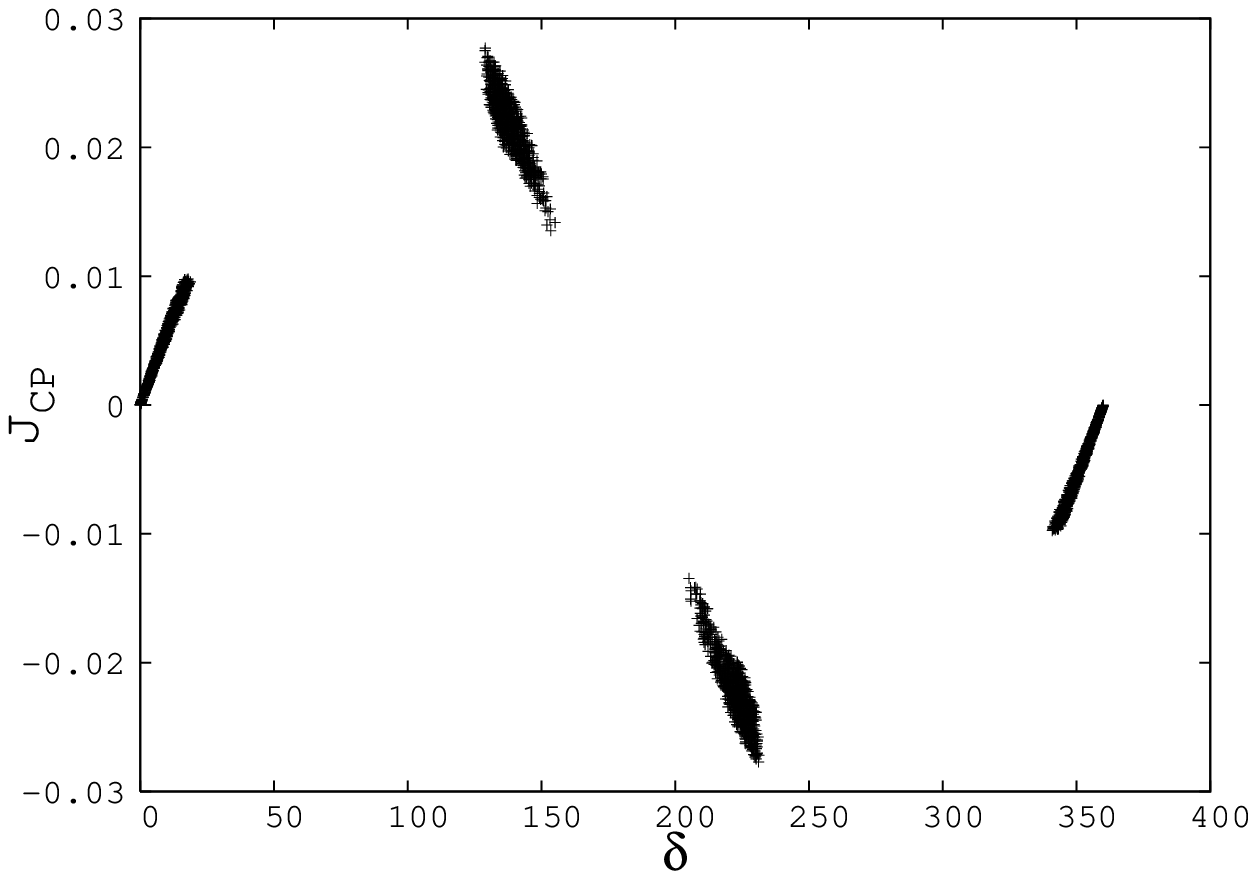}}
\caption{\label{fig2} Case $T_{2}$ (IO):  Scattering plots of  Majorana phases, Dirac CP-violating phase ($\delta$), effective neutrino mass  $|M|_{ee}$,  Jarlskog rephrasing invariant ($J_{CP}$) have been shown. All the phase angles ($\delta, \rho, \sigma$) are measured in degrees and $|M|_{ee}$ is in eV unit.  }
\end{center}
\end{figure}

Using Eqs. \ref{eq6}, \ref{eq9} and \ref{eq10}, we obtain the following analytical relations in the leading order approximation  of $\theta_{13}$
\begin{equation}\label{eq23}
\frac{\lambda_{1}}{\lambda_{3}}\approx \;- sec2 \theta_{12}( c_{12}^{2}-t^{2}_{23}e^{2i\delta}),
\end{equation}
\begin{equation}\label{eq24}
\frac{\lambda_{2}}{\lambda_{3}}\approx \; sec2 \theta_{12} (s_{12}^{2}-t^{2}_{23}e^{2i\delta}).
\end{equation}
From Eqs. \ref{eq23} and \ref{eq24}, one can obtain the neutrino mass ratios
\begin{equation}\label{eq25}
\xi \approx sec2 \theta_{12} \sqrt{c^{4}_{12}+t^{4}_{23}-2c^{2}_{12}t^{2}_{23}cos2 \delta},
\end{equation}
\begin{equation}\label{eq26}
\zeta \approx sec2 \theta_{12} \sqrt{s^{4}_{12}+t^{4}_{23}-2s^{2}_{12}t^{2}_{23}cos2 \delta},
\end{equation}
and the Majorana CP-violating phases
\begin{equation}\label{eq27}
\rho \approx \frac{1}{2}tan^{-1}\bigg(- \frac{t^{2}_{23}sin 2\delta} {c^{2}_{12}-t^{2}_{23}cos2\delta }\bigg) +O(s_{13}) ,
\end{equation}
\begin{equation}\label{eq28}
\sigma \approx \frac{1}{2}tan^{-1}\bigg(- \frac{t^{2}_{23}sin 2\delta} {s^{2}_{12}-t^{2}_{23}cos2\delta }\bigg) +O(s_{13}). 
\end{equation}
The correlation plots for case $T_{2}$ have been compiled in Fig. \ref{fig1} (a, b, c, d) and Fig. \ref{fig2} (a, b, c, d), respectively. It is found from the analysis that case $T_{2}$ favors both normal (NO) and inverted mass ordering (IO)  at 3$\sigma$ CL. The parameter space of CP violating phases $\delta, \rho, \sigma$ is found to be constrained to very small ranges for NO [Fig. \ref{fig1}(a, b)]. However, for IO, comparatively significant allowed parameter space is available for $\delta, \rho, \sigma$ [Fig. \ref{fig2}(a, b)]. \\
In the leading order approximation of $s_{13}$, the effective mass term in $0\nu \beta \beta $ decay turns out to be               
 \begin{equation}\label{eq29}
 |M|_{ee} \approx m_{3}t^{2}_{23} \approx 2.32 \times 10^{-2} eV,
 \end{equation}
 which lies well within the sensitivity limit of neutrinoless double beta decay experiments. Fig. \ref{fig1}(c) and Fig. \ref{fig2}(c)  show  the correlation plot between $|M|_{ee}$ and $\delta$ for NO and IO respectively.  The Jarlskog rephrasing parameter $J_{CP}$ is found to be non vanishing for NO [Fig. \ref{fig1}(d)], however, $J_{CP}=0$ cannot be excluded for IO [Fig. \ref{fig2}(d)].
  
 \subsection{Case $T_{3}$}
\begin{figure}[h!]
\begin{center}
\subfigure[]{\includegraphics[width=0.35\columnwidth]{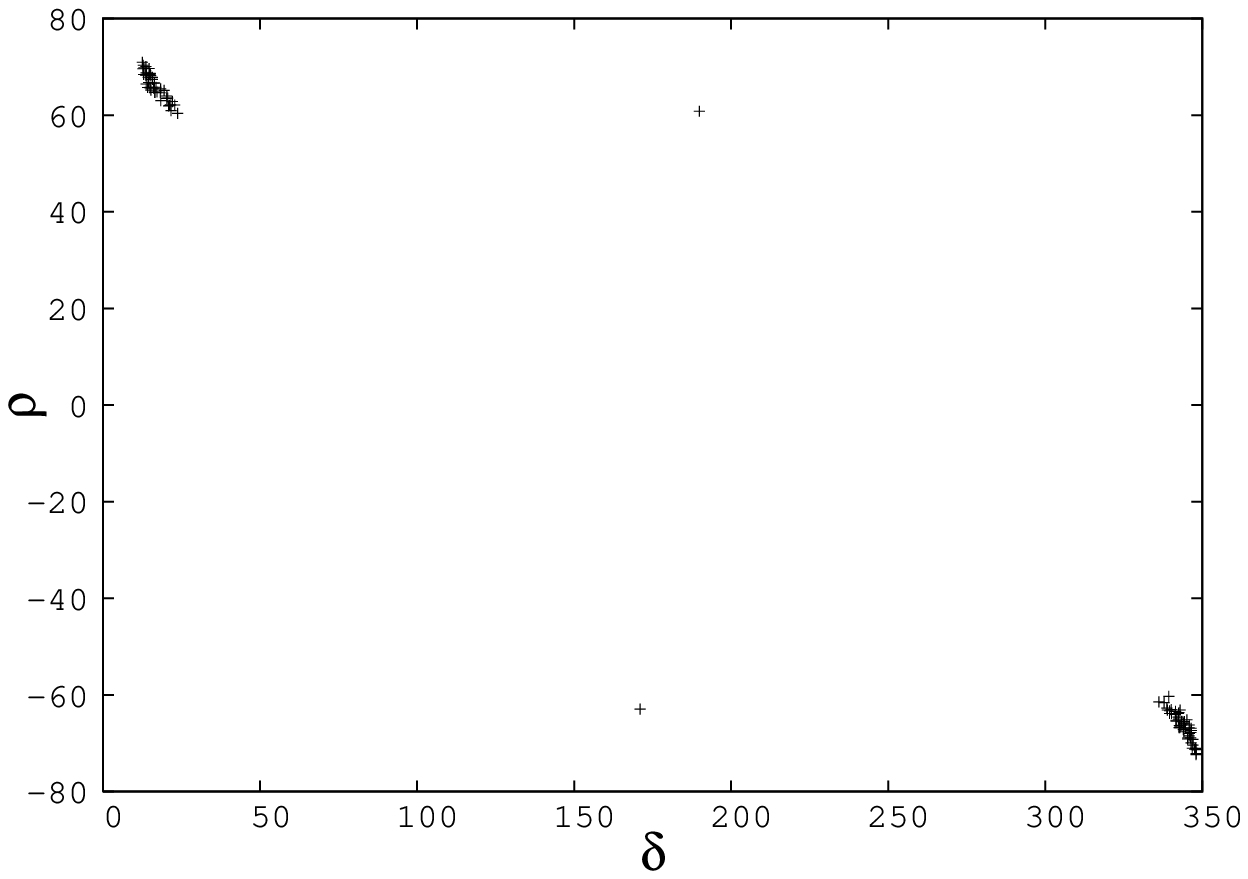}} \ \ \
\subfigure[]{\includegraphics[width=0.35\columnwidth]{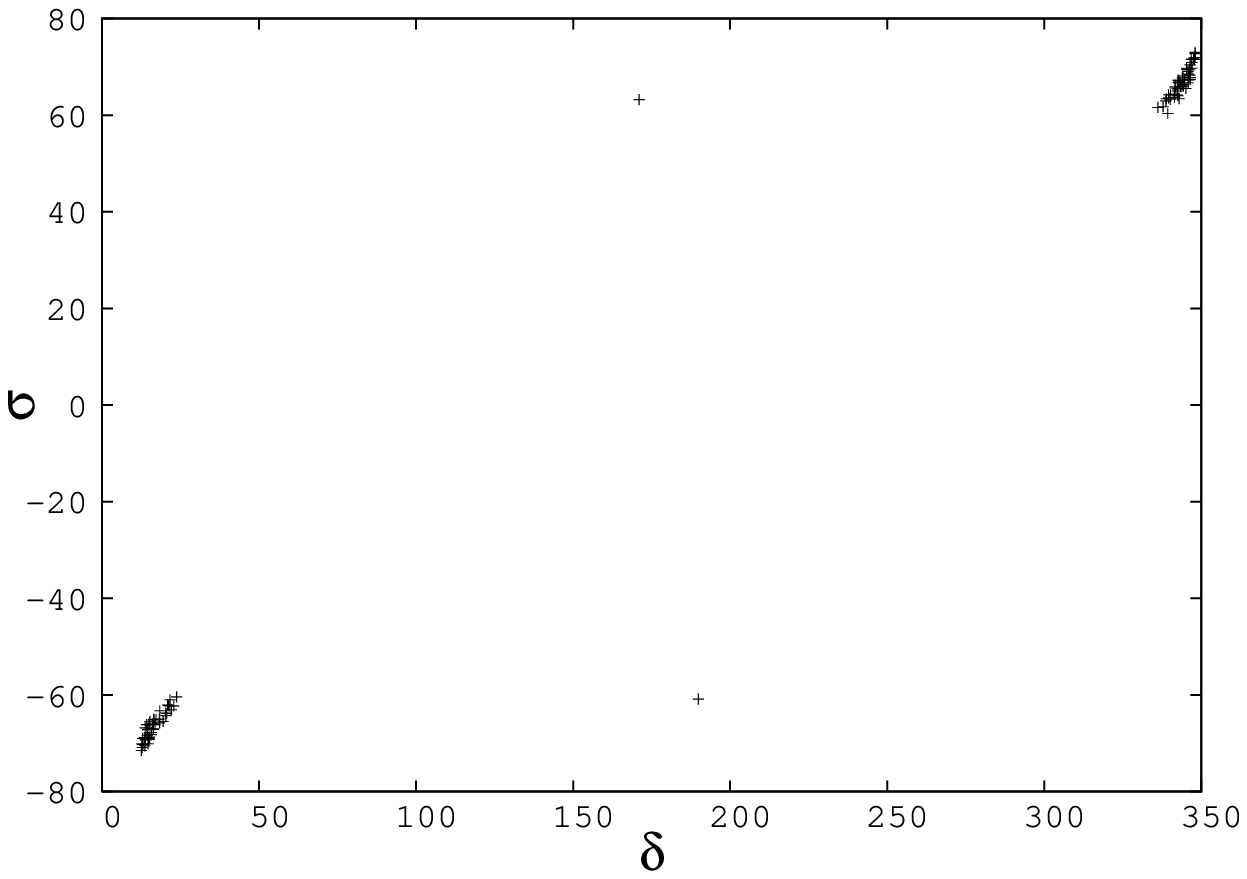}}\\
\subfigure[]{\includegraphics[width=0.35\columnwidth]{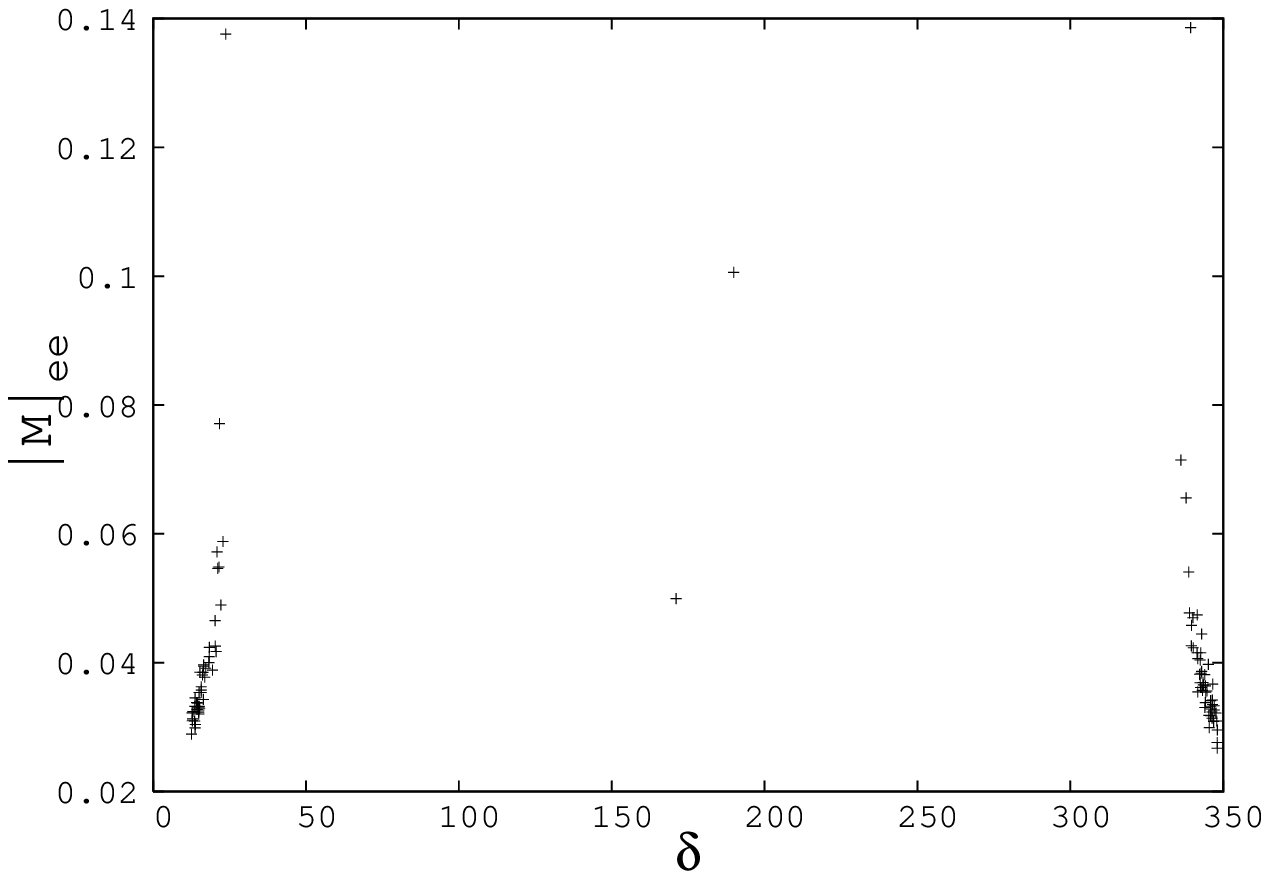}} \ \ \
\subfigure[]{\includegraphics[width=0.35\columnwidth]{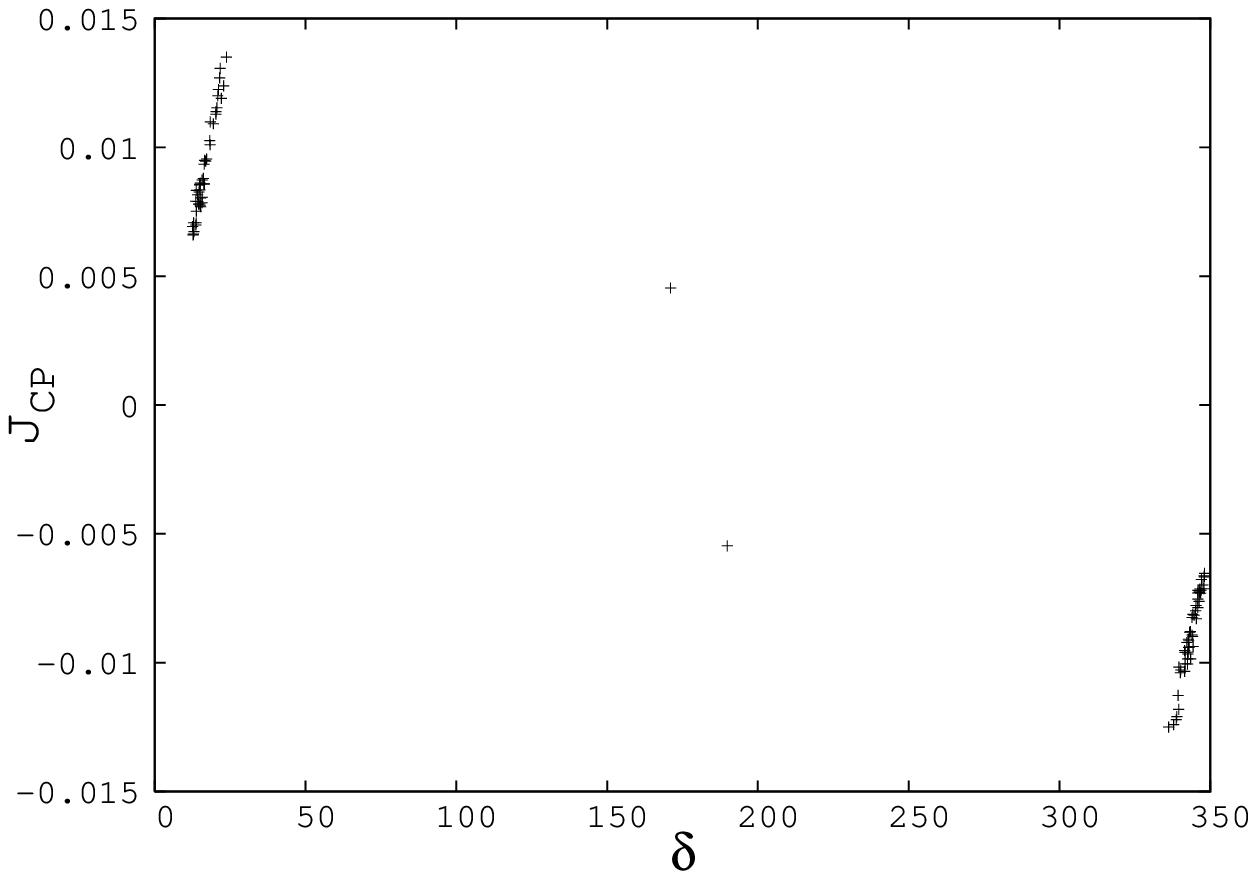}}
\caption{\label{fig3} Case $T_{3}$ (NO):  Scattering plots of  Majorana phases, Dirac CP-violating phase ($\delta$), effective neutrino mass  $|M|_{ee}$,  Jarlskog rephrasing invariant ($J_{CP}$) have been shown. All the phase angles ($\delta, \rho, \sigma$) are measured in degrees and $|M|_{ee}$ is in eV unit.  }
\end{center}
\end{figure}

\begin{figure}[h!]
\begin{center}
\subfigure[]{\includegraphics[width=0.35\columnwidth]{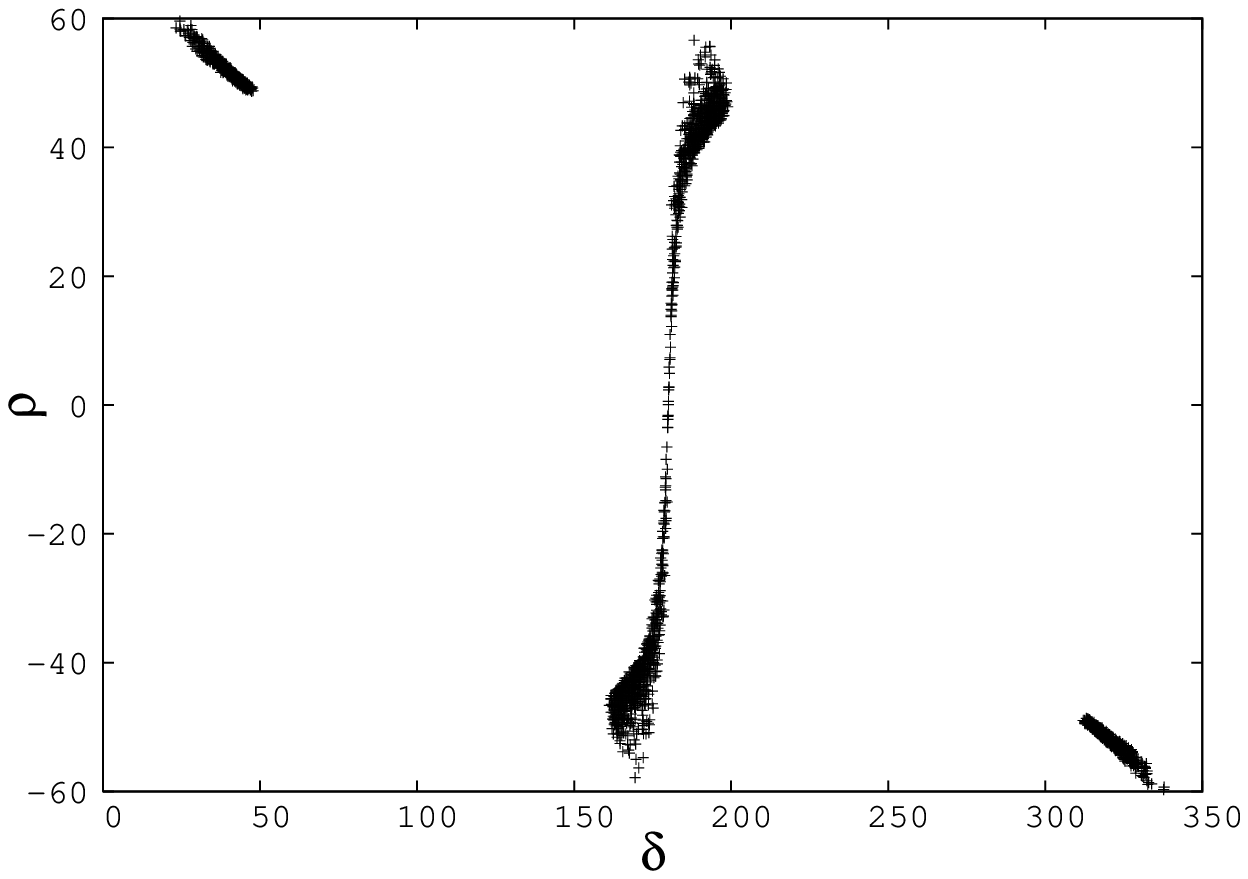}} \ \ \
\subfigure[]{\includegraphics[width=0.35\columnwidth]{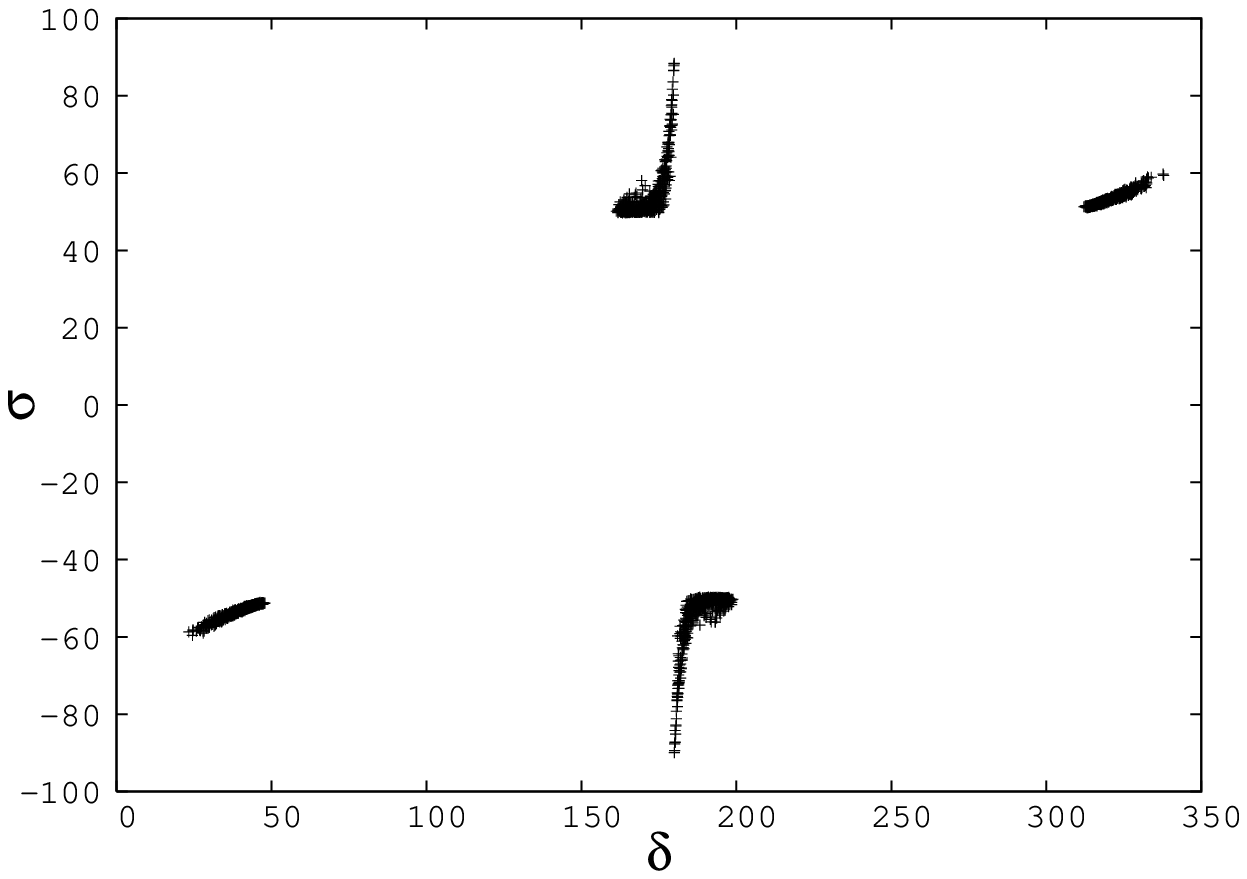}}\\
\subfigure[]{\includegraphics[width=0.35\columnwidth]{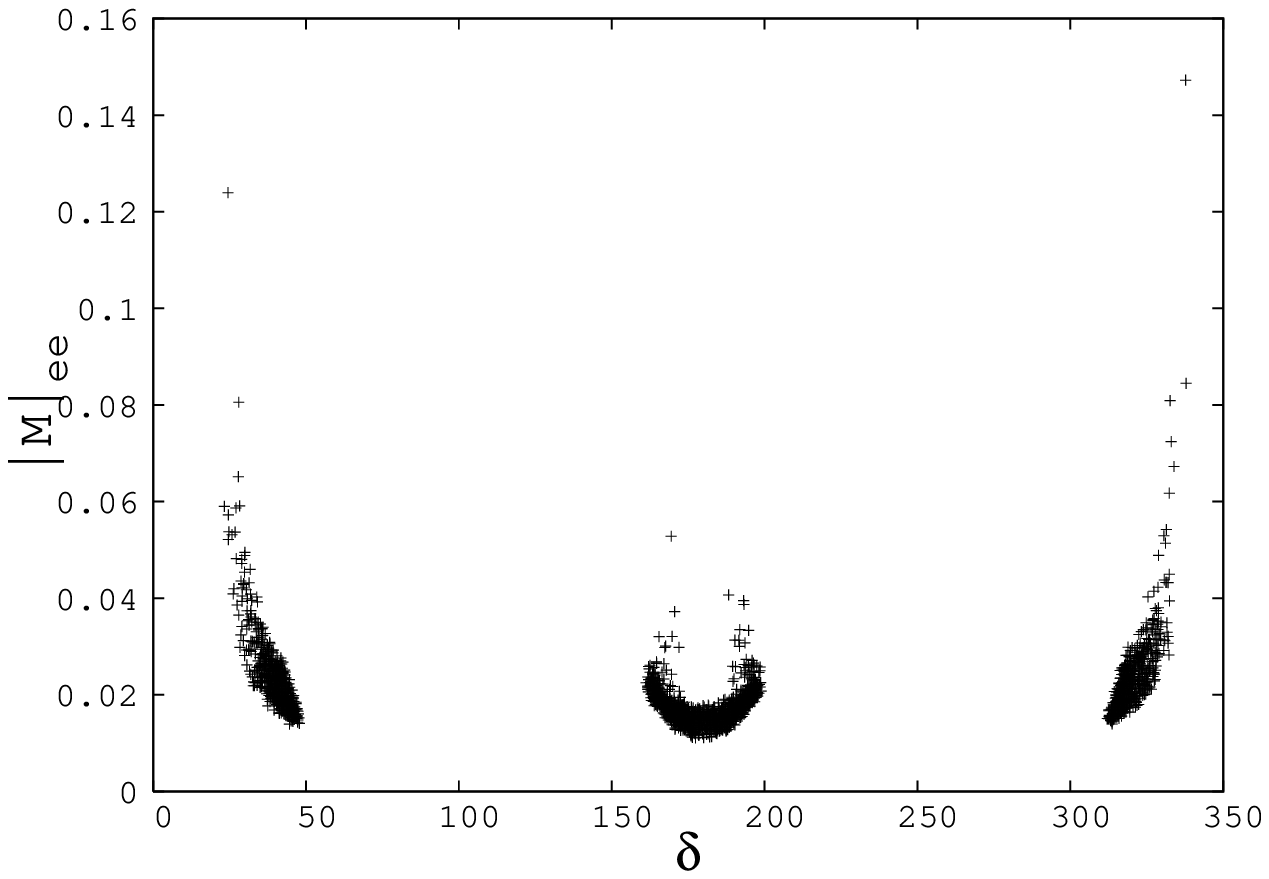}} \ \ \
\subfigure[]{\includegraphics[width=0.35\columnwidth]{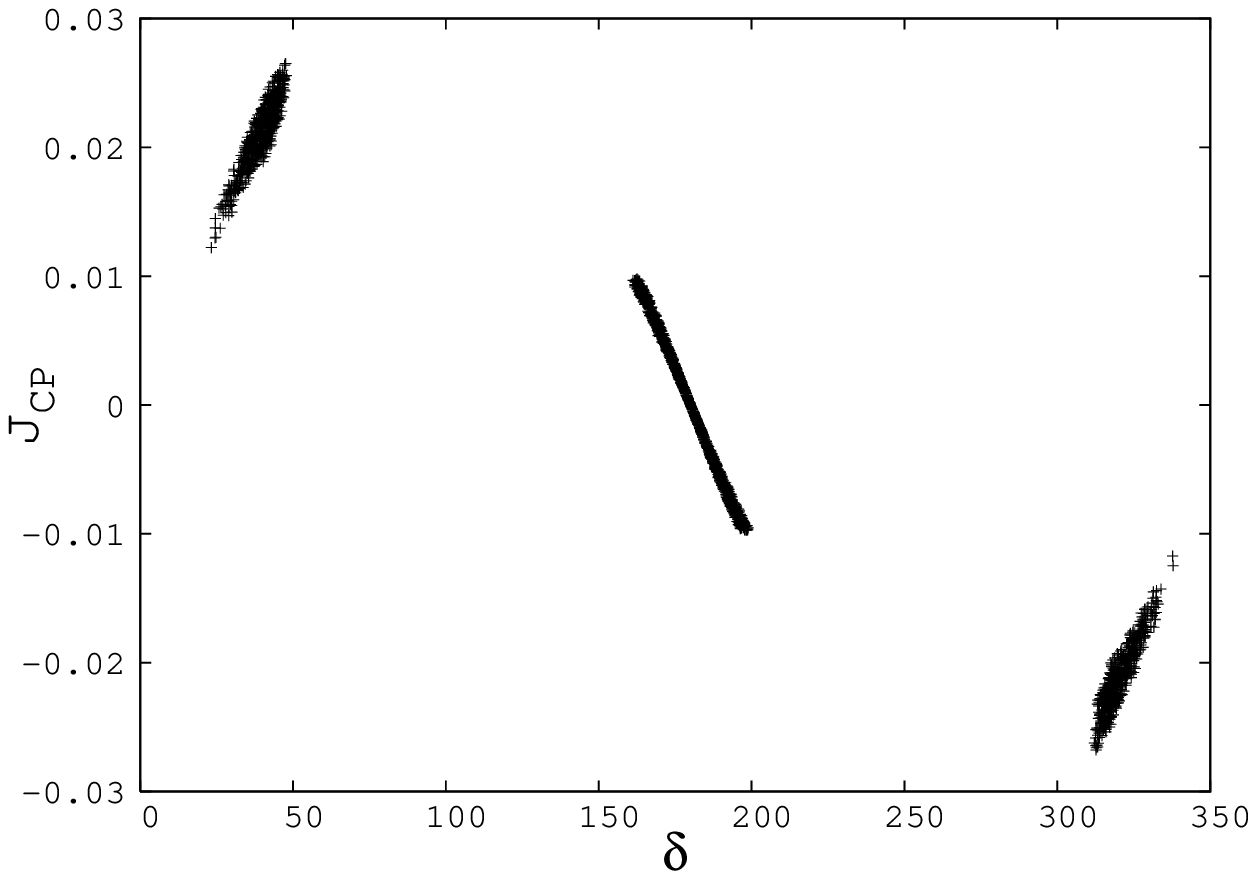}}
\caption{\label{fig4} Case $T_{3}$ (IO):  Scattering plots of  Majorana phases, Dirac CP-violating phase ($\delta$), effective neutrino mass  $|M|_{ee}$,  Jarlskog rephrasing invariant ($J_{CP}$) have been shown. All the phase angles ($\delta, \rho, \sigma$) are measured in degrees and $|M|_{ee}$ is in eV unit.  }
\end{center}
\end{figure}

 With the help of Eqs. \ref{eq6}, \ref{eq9} and \ref{eq10}, we deduce the following analytical expressions  in the leading order of $s_{13}$ term. 
 \begin{equation}\label{eq30}
\frac{\lambda_{1}}{\lambda_{3}}\approx \; sec2 \theta_{12}( c_{12}^{2}-\frac{1}{t^{2}_{23}}e^{2i\delta}),
\end{equation}
\begin{equation}\label{eq31}
\frac{\lambda_{2}}{\lambda_{3}}\approx \; sec2 \theta_{12} (s_{12}^{2}-\frac{1}{t^{2}_{23}}e^{2i\delta}).
\end{equation}
From the Eqs. \ref{eq30} and \ref{eq31}, one can obtain the neutrino mass ratios
\begin{equation}\label{eq32}
\xi \approx sec2 \theta_{12} \sqrt{c^{4}_{12}+\frac{1}{t^{4}_{23}}-2c^{2}_{12}\frac{1}{t^{2}_{23}}cos2 \delta},
\end{equation}
\begin{equation}\label{eq33}
\zeta \approx sec2 \theta_{12} \sqrt{s^{4}_{12}+\frac{1}{t^{4}_{23}}-2s^{2}_{12}\frac{1}{t^{2}_{23}}cos2 \delta},
\end{equation}
and  the Majorana CP violating phases
\begin{equation}\label{eq34}
\rho \approx \frac{1}{2}tan^{-1}\bigg(- \frac{sin 2\delta} {c^{2}_{12}t^{2}_{23}-cos2\delta }\bigg) +O(s_{13}), 
\end{equation}
\begin{equation}\label{eq35}
\sigma \approx \frac{1}{2}tan^{-1}\bigg(- \frac{sin 2\delta} {s^{2}_{12}t^{2}_{23}-t^{2}_{23}cos2\delta }\bigg) +O(s_{13}) .
\end{equation}
\\
Cases $T_{2}$ and $T_{3}$ are related via permutation symmetry, therefore the  phenomenological results for case $T_{3}$  can be obtained from case $T_{3}$ by using Eq. \ref{eq3}. The correlation plots for case $T_{3}$ have been complied in Fig. \ref{fig3}(a, b, c, d) (NO) and Fig. \ref{fig4}(a, b, c, d) (IO),  indicating the parameter space of  $\rho, \sigma, \delta$, $|M|_{ee}$, $J_{CP}$.


 \subsection{Case $T_{4}$}
 Using Eqs. \ref{eq6}, \ref{eq9} and \ref{eq10}, we deduce the following analytical expressions in the leading order of $s_{13}$ term
 \begin{equation}\label{eq36}
\frac{ \lambda_{1}}{ \lambda_{3}}\approx \frac{ \lambda_{2}}{ \lambda_{3}}\approx -0.5.  
 \end{equation}
 \begin{equation}\label{eq37}
\xi \approx \zeta \approx 0.5.  
 \end{equation}
 Since $R_{\nu}=0$ in the leading order approximation of $s_{13}$, we have to work at next to leading order,
 and we get
 \begin{equation}\label{eq38}
 \frac{ \lambda_{1}}{\lambda_{3}} \approx -\;\frac{1}{2}\bigg(1-\frac{3}{2} \frac{s_{23} s_{13}}{c_{12}s_{12}c_{23}}e^{i\delta}\bigg)+O(s^{2}_{13}),
\end{equation}
\begin{equation}\label{eq39}
\frac{ \lambda_{2}}{\lambda_{3}} \approx -\;\frac{1}{2}\bigg(1+ \frac{3}{2} \frac{s_{23} s_{13}}{c_{12}s_{12}c_{23}}e^{i\delta}\bigg)+O(s^{2}_{13}).
\end{equation}
\begin{figure}[h!]
\begin{center}
\subfigure[]{\includegraphics[width=0.35\columnwidth]{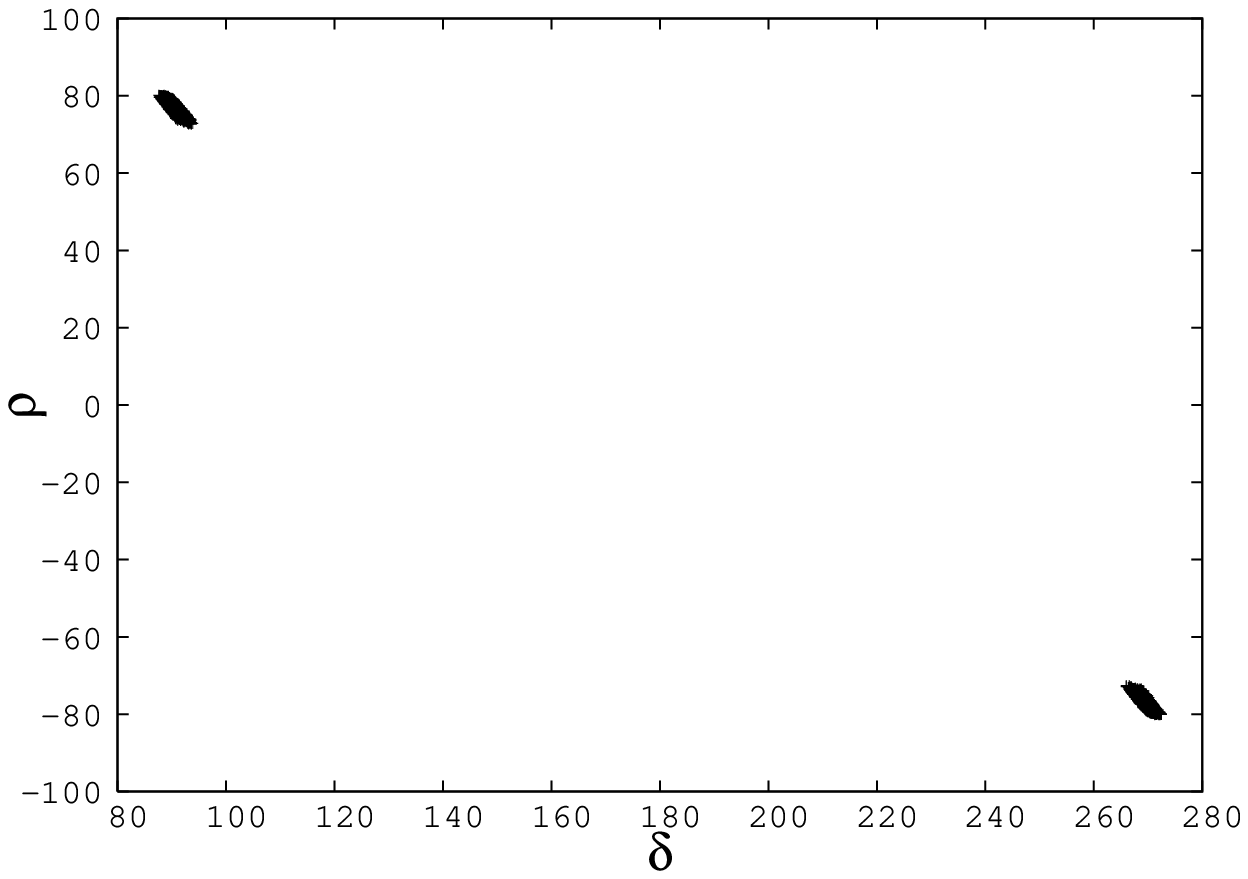}} \ \ \
\subfigure[]{\includegraphics[width=0.35\columnwidth]{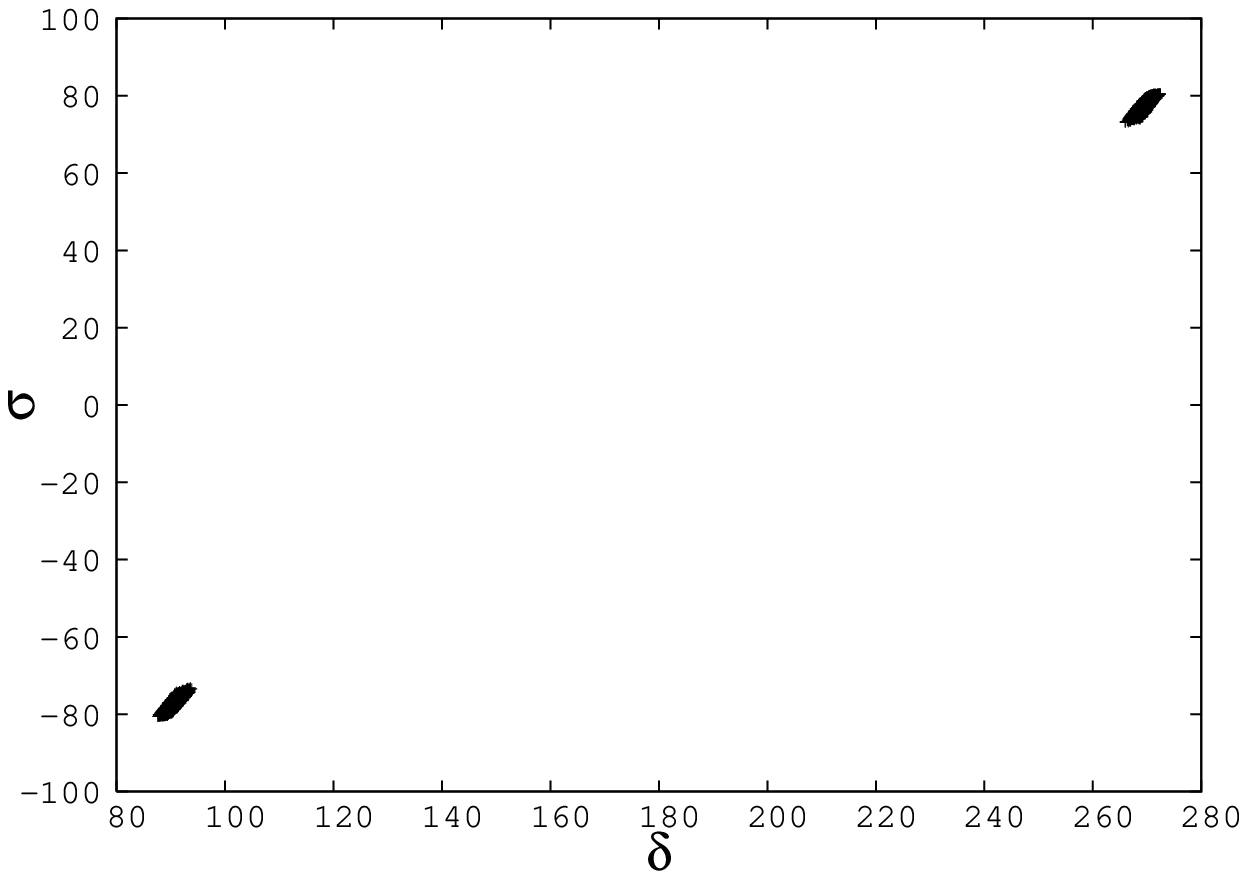}}\\
\subfigure[]{\includegraphics[width=0.35\columnwidth]{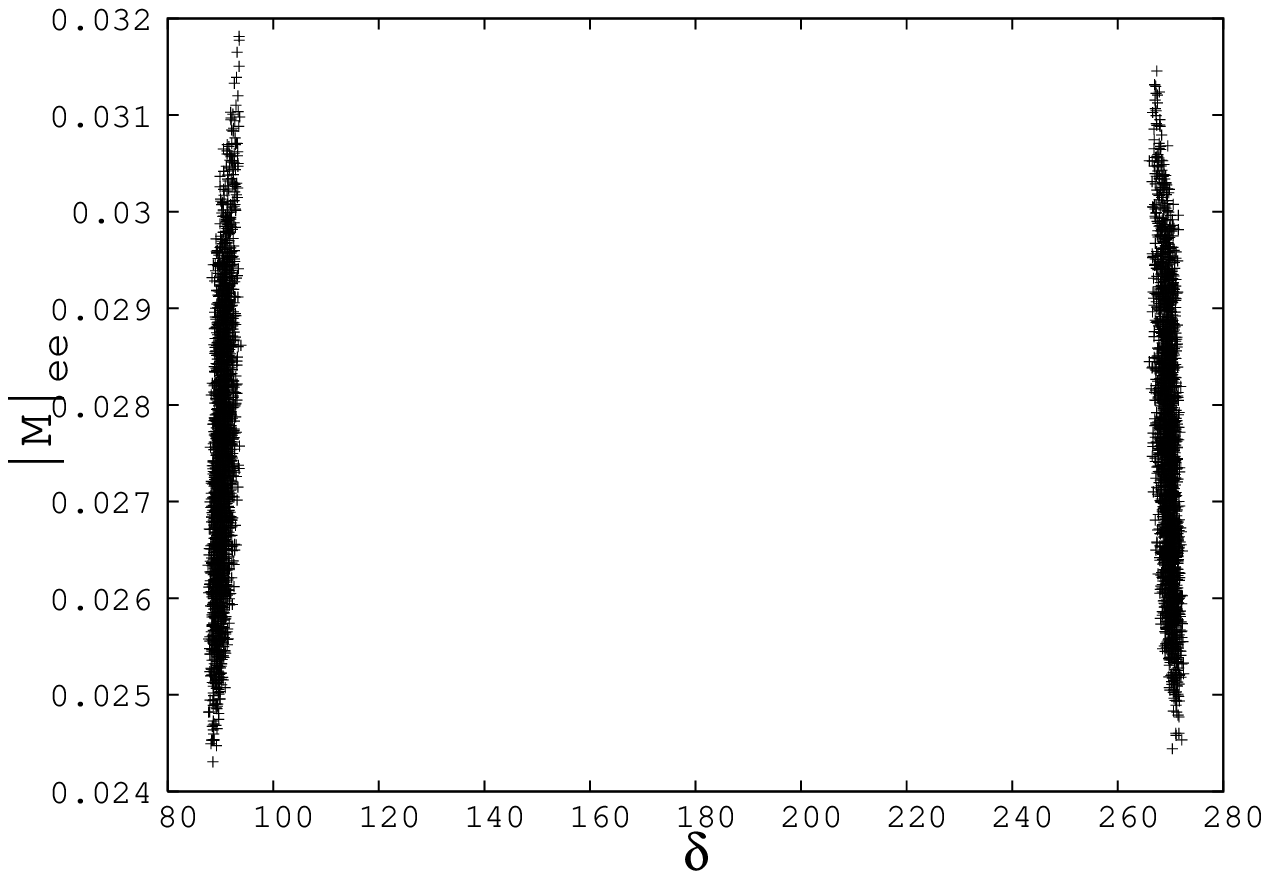}} \ \ \
\subfigure[]{\includegraphics[width=0.35\columnwidth]{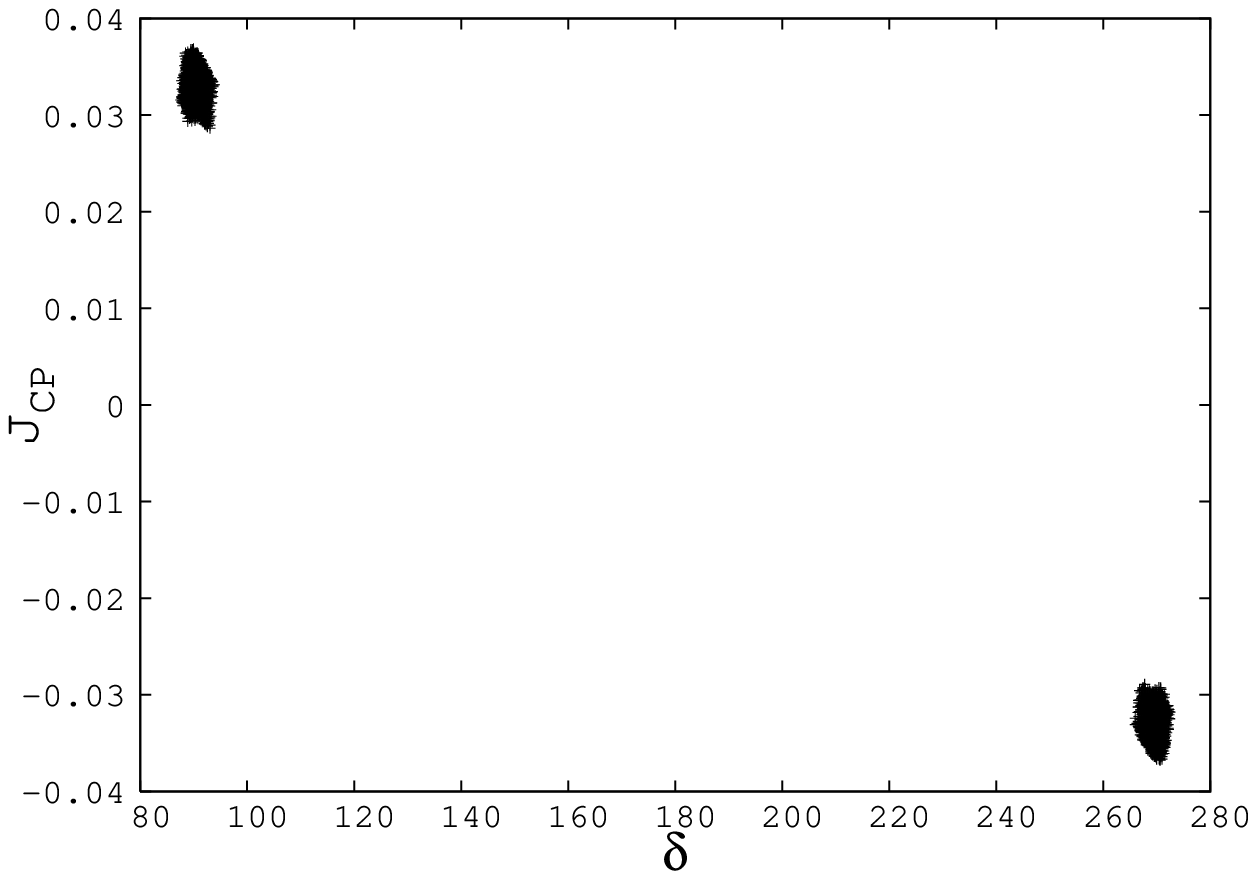}}
\caption{\label{fig5} Case $T_{4}$ (NO):  Scattering plots of  Majorana phases, Dirac CP-violating phase ($\delta$), effective neutrino mass  $|M|_{ee}$,  Jarlskog rephrasing invariant ($J_{CP}$) have been shown. All the phase angles ($\delta, \rho, \sigma$) are measured in degrees and $|M|_{ee}$ is in eV unit.  }
\end{center}
\end{figure}
 Using Eqs. \ref{eq38} and \ref{eq39}, the neutrino mass ratios can be given as 
 
 \begin{equation}\label{eq40}
 \xi \approx \frac{1}{2} \sqrt{1+\frac{9}{4}\frac{s_{23}^{2}s_{13}^{2}}{c_{12}^{2}s_{12}^{2}c_{23}^{2}}-\frac{3s_{23}s_{13}}{c_{12}s_{12}c_{23}}cos\delta},
 \end{equation}
 \begin{equation}\label{eq41}
 \zeta \approx \frac{1}{2} \sqrt{1+\frac{9}{4}\frac{s_{23}^{2}s_{13}^{2}}{c_{12}^{2}s_{12}^{2}c_{23}^{2}}+\frac{3s_{23}s_{13}}{c_{12}s_{12}c_{23}}cos\delta},
 \end{equation}
 and the Majorana CP violating phases as
 \begin{equation}\label{eq42}
 \rho \approx \frac{1}{2}tan^{-1} \bigg( \frac{3s_{23}s_{13}sin\delta} {3s_{23}s_{13}cos\delta-2c_{12}s_{12}c_{23}}\bigg)+O(s_{13}^{2}),
 \end{equation}
 \begin{equation}\label{eq43}
 \sigma \approx \frac{1}{2}tan^{-1} \bigg( \frac{3s_{23}s_{13}sin\delta} {3s_{23}s_{13}cos\delta+2c_{12}s_{12}c_{23}}\bigg)+O(s_{13}^{2}).
 \end{equation}
 Using the best fit values from latest global fits on neutrino oscillation data [Table \ref{tab1}], the neutrino mass spectrum can be given as follows
\begin{equation}\label{eq44}
m_{3}= \sqrt{\frac{\delta m^{2}}{(\zeta^{2}-\xi^{2})}}\approx 4.87\times 10^{-2}eV,
\end{equation}
\begin{equation}\label{eq45}
 m_{2}= m_{3}\zeta \approx 2.85\times 10^{-2}eV,
\end{equation}
 \begin{equation}\label{eq46}
  m_{1}=m_{3}\xi\approx 2.71\times 10^{-2}eV, 
 \end{equation}
implying that only NO is allowed. Fig. \ref{fig5}(a, b) show the correlation plot between Majorana phases ($\rho,
\sigma$) and Dirac CP violating phase ($\delta$). The parameter space for $\delta$ is found to be restricted near  $90^{0}$ and $270^{0}$. The prediction is significant considering the latest hint on $\delta$ near $270^{0}$ in the recent global fits on neutrino oscillation data [Table \ref{tab1}]. The Majorana phases ($\rho, \sigma$) are found to be constrained near $-90^{0}$ and $90^{0}$. In Fig. \ref{fig5}(d), it is explicitly shown that $J_{CP}$ is non-zero implying that Case $T_{5}$ is necessarily CP violating.  \\
 In the leading order of $s_{13}$, the effective mass term in $0\nu \beta \beta $ decay  can be  approximated as               
 \begin{equation}\label{eq47}
 |M|_{ee} \approx \frac{m_{3}}{2} \approx 2.43 \times 10^{-2} eV,
 \end{equation}
which is well within the accessible limit of next generation neutrinoless double decay experiments. The correlation plot between 
$|M|_{ee}$ and $\delta$ has been provided for case $T_{4}$ in Fig. \ref{fig5}(c).
 
 \subsection{Case $T_{5}$}
\begin{figure}[h!]
\begin{center}
\subfigure[]{\includegraphics[width=0.35\columnwidth]{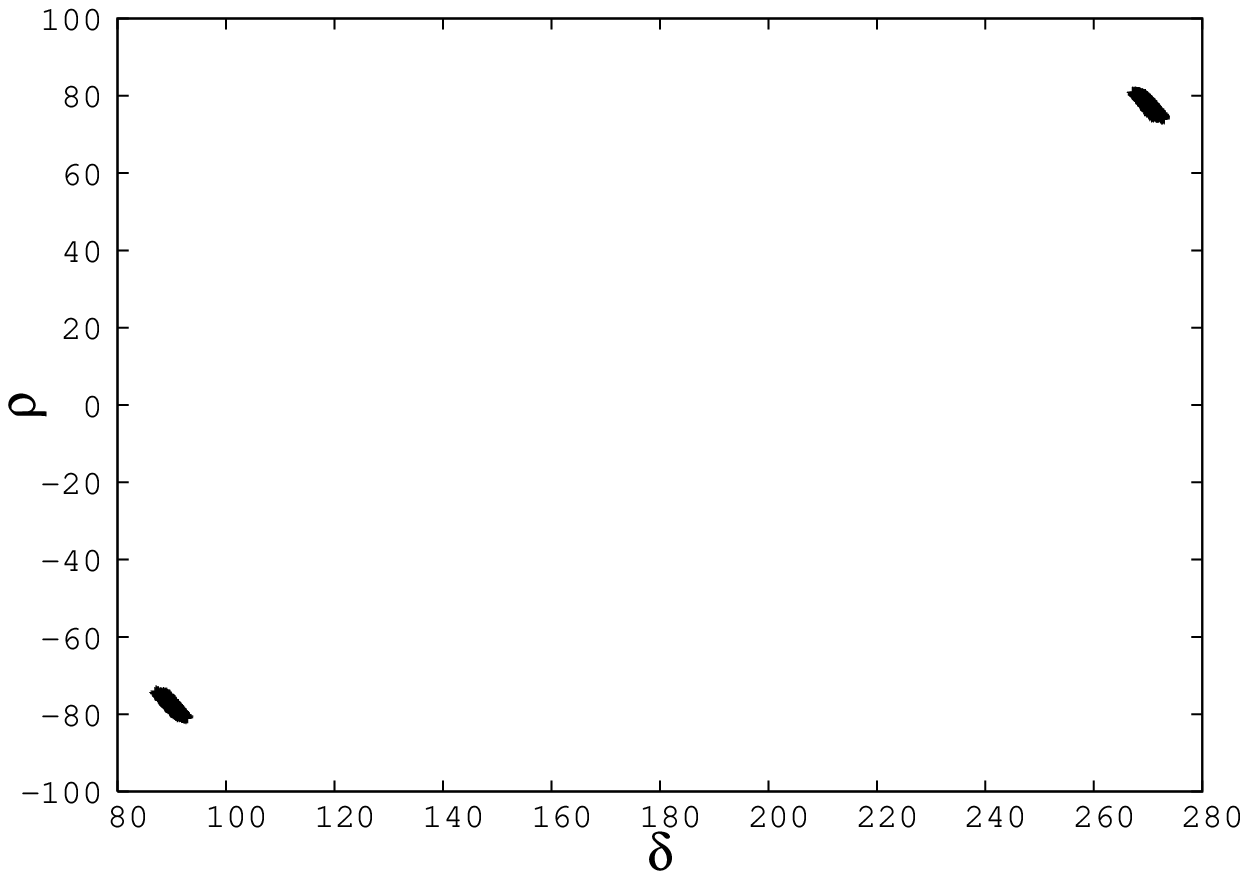}} \ \ \
\subfigure[]{\includegraphics[width=0.35\columnwidth]{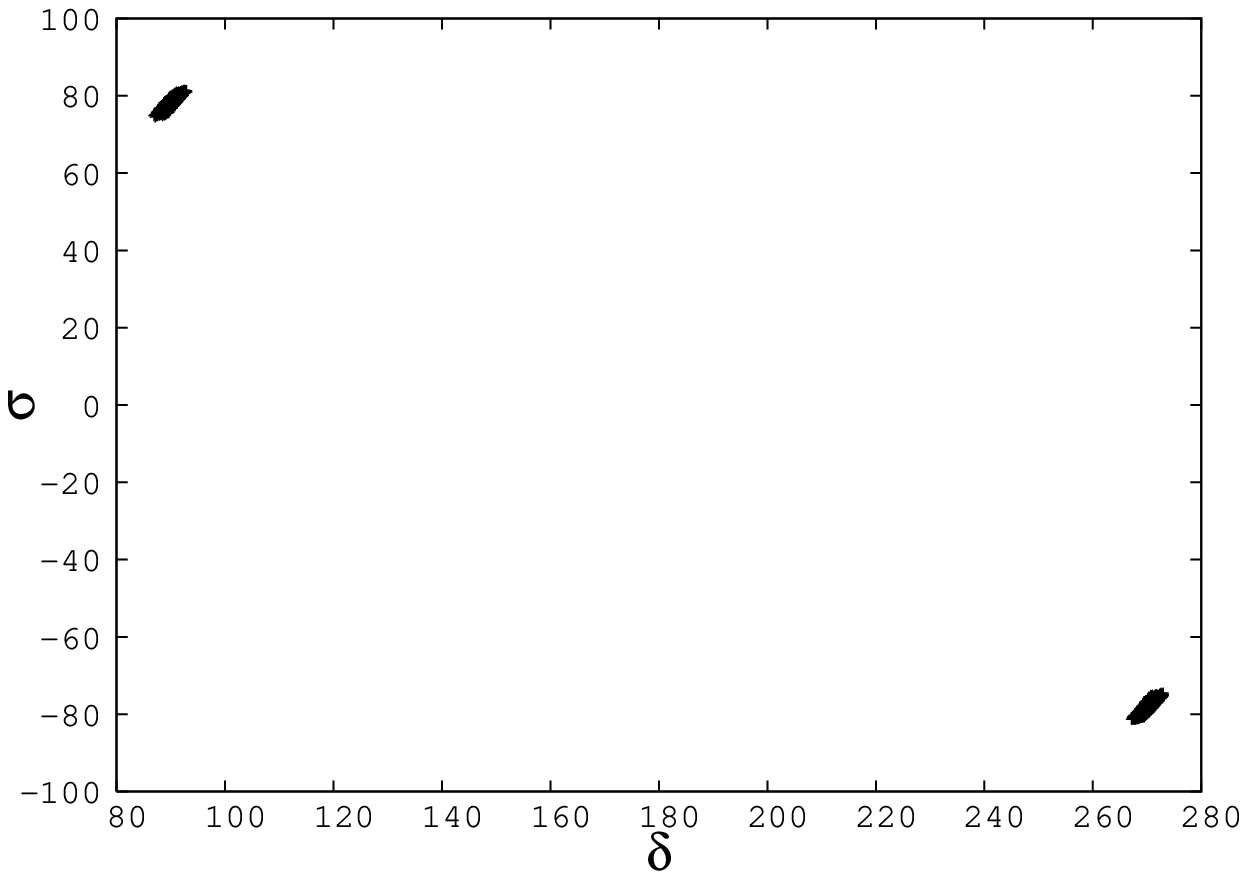}}\\
\subfigure[]{\includegraphics[width=0.35\columnwidth]{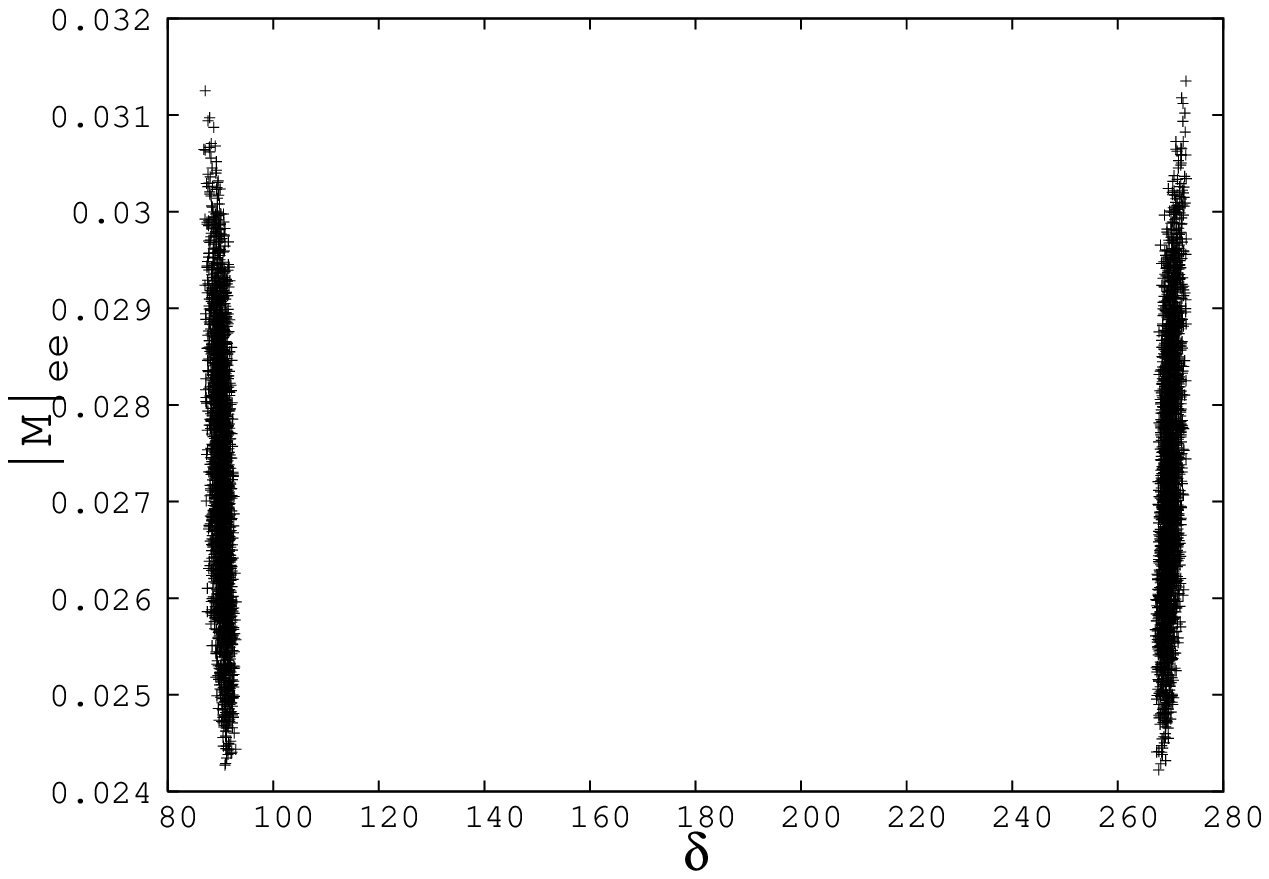}} \ \ \
\subfigure[]{\includegraphics[width=0.35\columnwidth]{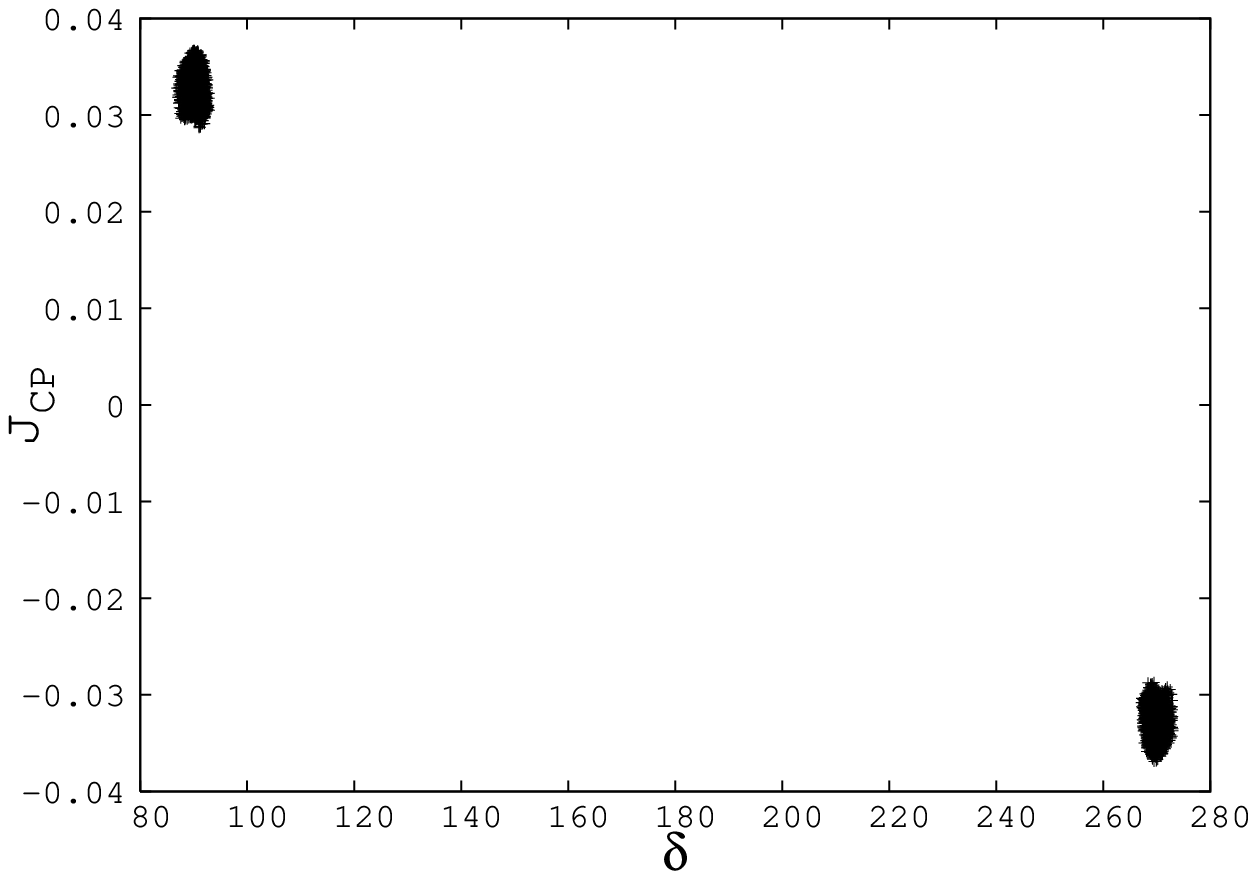}}
\caption{\label{fig6} Case $T_{5}$ (NO):  Scattering plots of  Majorana phases, Dirac CP-violating phase ($\delta$), effective neutrino mass  $|M|_{ee}$,  Jarlskog rephrasing invariant ($J_{CP}$) have been shown. All the phase angles ($\delta, \rho, \sigma$) are measured in degrees and $|M|_{ee}$ is in eV unit.  }
\end{center}
\end{figure}

 With the help of Eqs. \ref{eq6}, \ref{eq9} and \ref{eq10}, we obtain the following analytical expressions in the leading order of $s_{13}$ term
 \begin{equation}\label{eq48}
\frac{ \lambda_{1}}{ \lambda_{3}}\approx \frac{ \lambda_{2}}{ \lambda_{3}}\approx 0.5,  
 \end{equation}
 \begin{equation}\label{eq49}
\xi \approx \zeta \approx 0.5.  
 \end{equation}
 Since $R_{\nu}=0$ in the leading order approximation of $s_{13}$, we have to work at next to leading order,  and we obtain
 \begin{equation}\label{eq50}
 \frac{ \lambda_{1}}{\lambda_{3}} \approx -\;\frac{1}{2}\bigg(1+\frac{3}{2} \frac{c_{23} s_{13}}{c_{12}s_{12}s_{23}}e^{i\delta}\bigg)+O(s^{2}_{13}),
\end{equation}
\begin{equation}\label{eq51}
\frac{ \lambda_{2}}{\lambda_{3}} \approx -\;\frac{1}{2}\bigg(1-\frac{3}{2} \frac{c_{23} s_{13}}{c_{12}s_{12}s_{23}}e^{i\delta}\bigg)+O(s^{2}_{13}).
\end{equation}
 From Eqs. \ref{eq50} and \ref{eq51}, the neutrino mass ratios can be given as follow
 \begin{equation}\label{eq52}
 \xi \approx \frac{1}{2} \sqrt{1+\frac{9}{4}\frac{c_{23}^{2}s_{13}^{2}}{c_{12}^{2}s_{12}^{2}s_{23}^{2}}-\frac{3c_{23}s_{13}}{c_{12}s_{12}s_{23}}cos\delta},
 \end{equation}
 \begin{equation}\label{eq53}
 \zeta \approx \frac{1}{2} \sqrt{1+\frac{9}{4}\frac{c_{23}^{2}s_{13}^{2}}{c_{12}^{2}s_{12}^{2}s_{23}^{2}}+\frac{3c_{23}s_{13}}{c_{12}s_{12}s_{23}}cos\delta},
 \end{equation}
 and the Majorana CP violating phases as
 \begin{equation}\label{eq54}
 \rho \approx \frac{1}{2}tan^{-1} \bigg( \frac{3c_{23}s_{13}sin\delta} {3c_{23}s_{13}cos\delta+2c_{12}s_{12}s_{23}}\bigg)+O(s_{13}^{2}),
 \end{equation}
 \begin{equation}\label{eq55}
 \sigma \approx \frac{1}{2}tan^{-1} \bigg( \frac{3c_{23}s_{13}sin\delta} {3c_{23}s_{13}cos\delta-2c_{12}s_{12}s_{23}}\bigg)+O(s_{13}^{2}).
 \end{equation}
Using the best fits from latest global neutrino oscillation data, the neutrino mass spectrum can be given as follows
\begin{equation}\label{eq56}
m_{3}= \sqrt{\frac{\delta m^{2}}{(\zeta^{2}-\xi^{2})}}\approx 4.87\times 10^{-2}eV,
\end{equation}
\begin{equation}\label{eq57}
 m_{2}= m_{3}\zeta \approx 2.85\times 10^{-2}eV,
\end{equation}
 \begin{equation}\label{eq58}
  m_{1}=m_{3}\xi\approx 2.71\times 10^{-2}eV, 
 \end{equation}
 indicating that only NO is allowed.  Since $T_{4}$ and $T_{5}$ are related due to permutation symmetry, therefore their phenomenological implications are similar. The phenomenological results for case $T_{5}$ can be derived from case $T_{4}$ using Eq. \ref{eq3}. The correlation plots for  $\rho, \sigma, \delta$, $|M|_{ee}$, $J_{CP}$ have been complied in Fig. \ref{fig6}.  \\
  In the leading order of $s_{13}$ term, the effective mass term in $0\nu \beta \beta $ decay can be approximated as
 \begin{equation}\label{eq59}
 |M|_{ee} \approx \frac{m_{3}}{2} \approx 2.43 \times 10^{-2} eV,
 \end{equation}
 which lies within the sensitivity limits of future $0\nu \beta \beta $ decay experiments. 
\subsection{Case $T_{6}$} 
With the help of Eqs. \ref{eq6}, \ref{eq9} and \ref{eq10}, we deduce the following analytical expressions in the leading order of $ s_{13}$ term
\begin{equation}\label{eq60}
\frac{\lambda_{1}}{\lambda_{3}}\approx -sec2\theta_{12} (c_{12}^{2}+e^{2i\delta})e^{i\delta},
\end{equation}
\begin{equation}\label{eq61}
\frac{\lambda_{2}}{\lambda_{3}}\approx sec2\theta_{12} (s_{12}^{2}+e^{2i\delta})e^{i\delta}.
\end{equation}
Using Eqs. \ref{eq60} and \ref{eq61}, we obtain the approximate relations for neutrino mass ratios
\begin{equation}\label{eq62}
\xi \approx sec2 \theta_{12} \sqrt{c_{12}^{4}+2c^{2}_{12}cos2 \delta+1},
\end{equation}
 \begin{equation}\label{eq63}
\zeta \approx sec2 \theta_{12} \sqrt{s_{12}^{4}+2s^{2}_{12}cos2\delta+1},
\end{equation}
 and the Majorana CP-violating phases
 \begin{equation}\label{eq64}
 \rho \approx \frac{1}{2} tan^{-1} \bigg(\frac{sin\delta (cos2\delta+c^{2}_{12})+sin2\delta cos \delta}{cos \delta (cos2\delta+c^{2}_{12})-sin 2\delta sin\delta}\bigg)+O(s_{13}^{2}),
 \end{equation}
 \begin{equation}\label{eq65}
 \sigma \approx \frac{1}{2} tan^{-1} \bigg(\frac{sin\delta (cos2\delta+s^{2}_{12})+sin2\delta cos \delta}{cos \delta (cos2\delta+s^{2}_{12})-sin 2\delta sin\delta}\bigg)+O(s_{13}^{2}).
 \end{equation}
 
\begin{figure}[h!]
\begin{center}
\subfigure[]{\includegraphics[width=0.35\columnwidth]{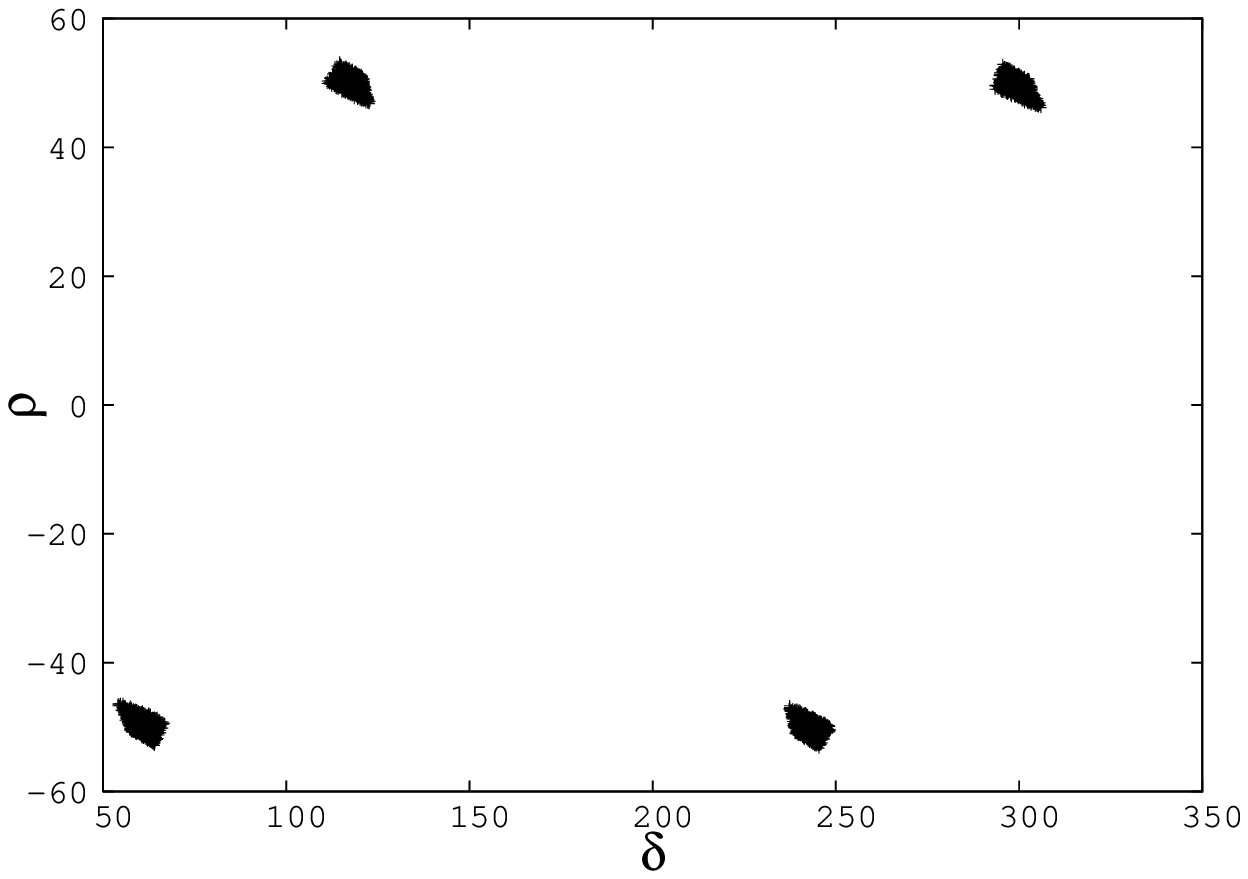}} \ \ \
\subfigure[]{\includegraphics[width=0.35\columnwidth]{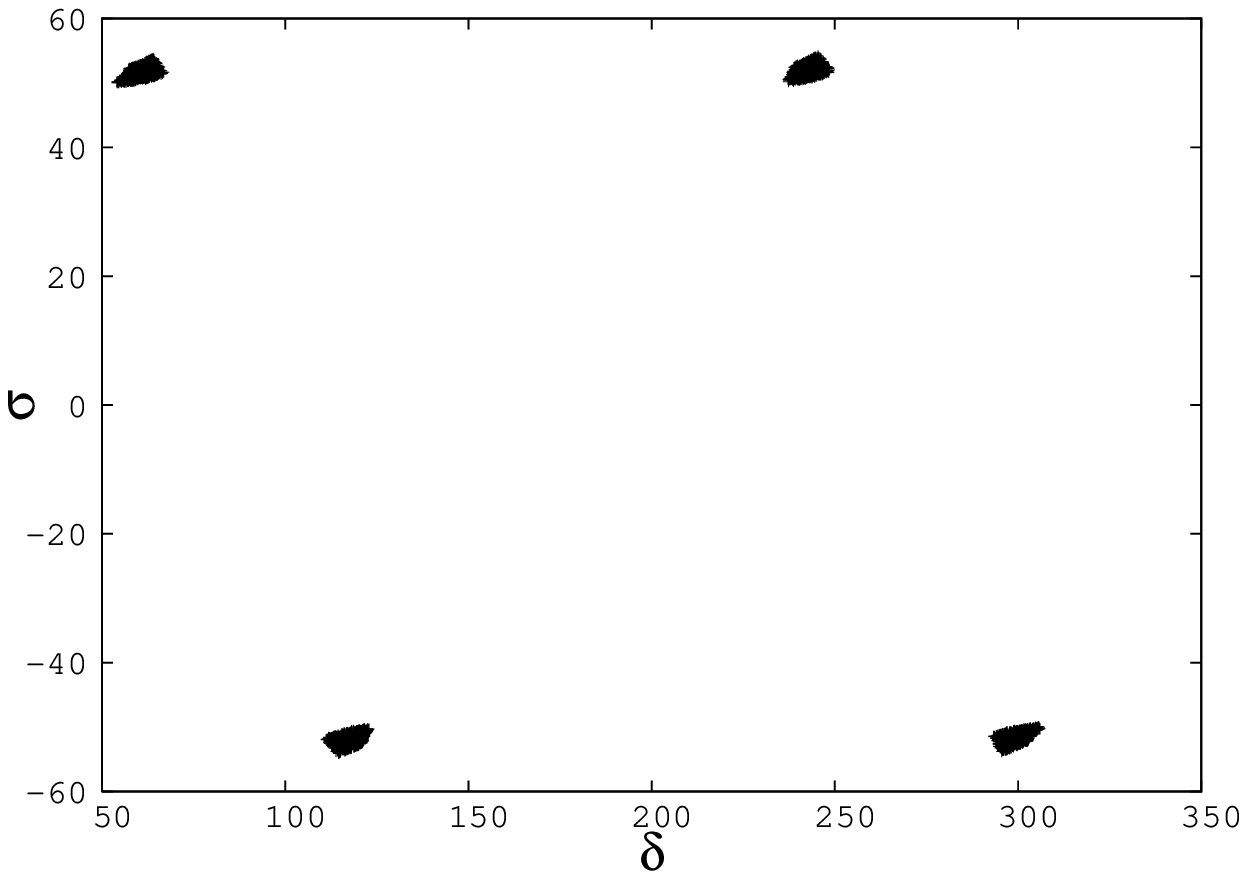}}\\
\subfigure[]{\includegraphics[width=0.35\columnwidth]{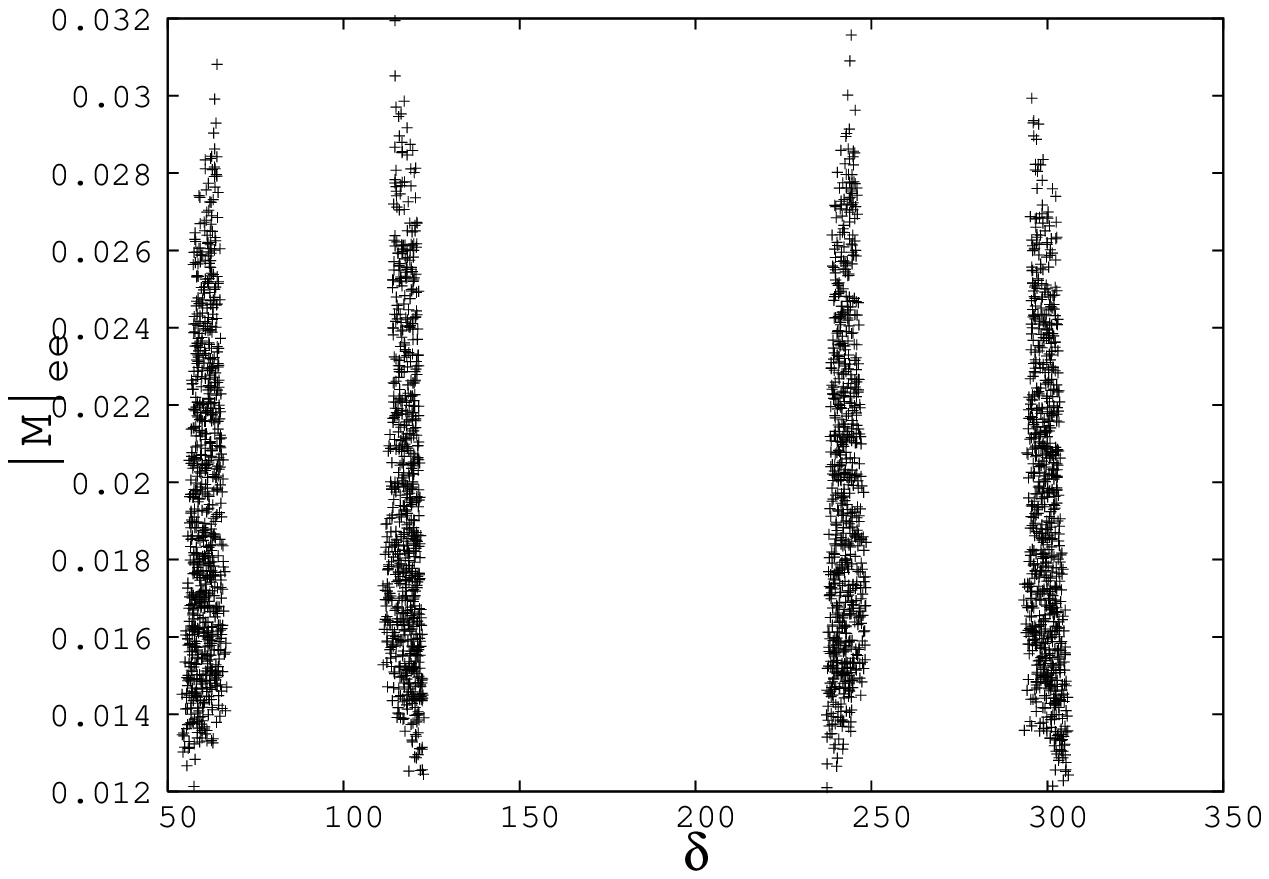}} \ \ \
\subfigure[]{\includegraphics[width=0.35\columnwidth]{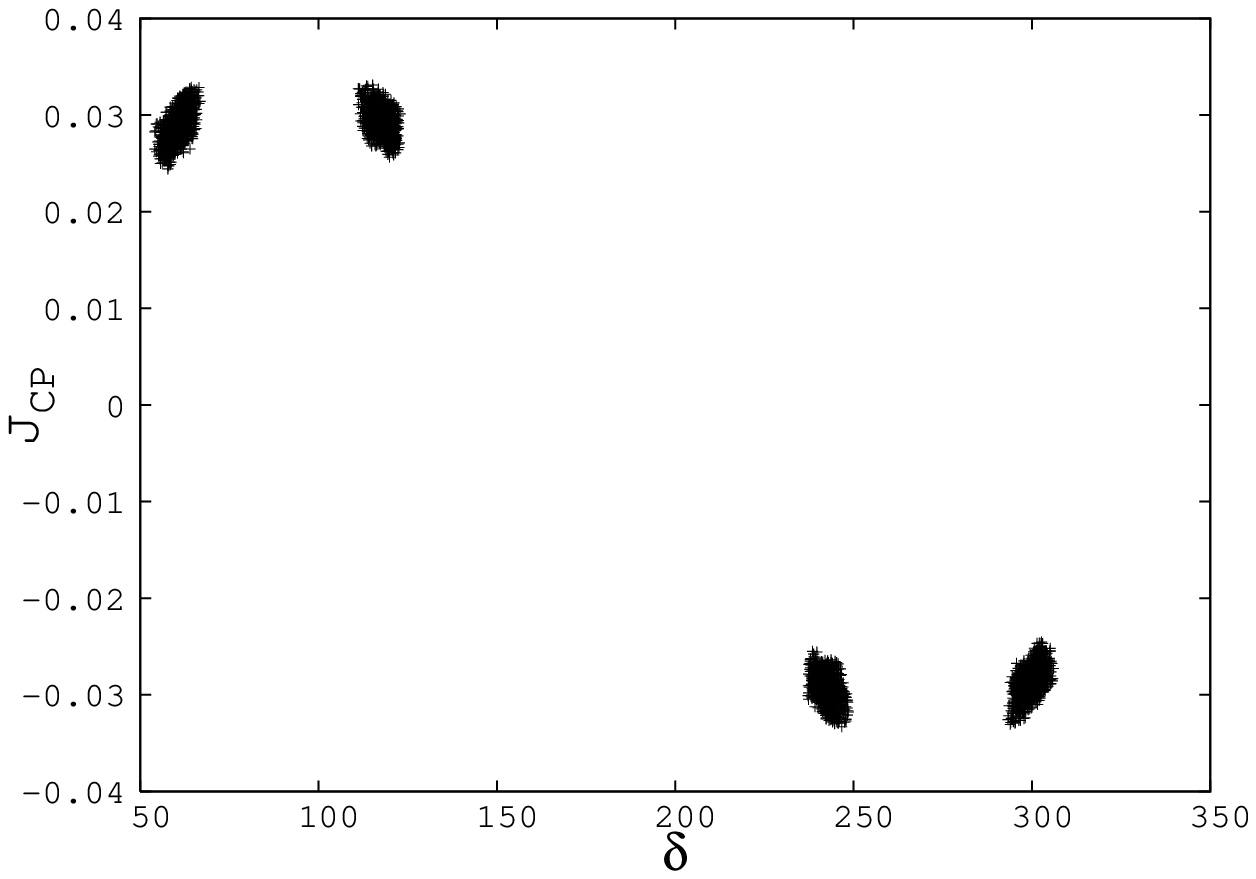}}
\caption{\label{fig7} Case $T_{6}$ (IO):  Scattering plots of  Majorana phases, Dirac CP-violating phase ($\delta$), effective neutrino mass  $|M|_{ee}$,  Jarlskog rephrasing invariant ($J_{CP}$) have been shown. All the phase angles ($\delta, \rho, \sigma$) are measured in degrees and $|M|_{ee}$ is in eV unit.  }
\end{center}
\end{figure}

The correlation plots for  $\rho, \sigma, \delta$, $|M|_{ee}$, $J_{CP}$ have been complied in Fig. \ref{fig7}(a,b,c,d).  In Fig. \ref{fig7}(a, b)  the parameter space of $\delta$ is found to be confined to very small regions. Also the Majorana phases $\rho$ and $\sigma $ also get restricted to  $-54.7- -44.7^{0} \oplus 45.3-54.5$. However, $J_{CP}=0$ as evident in Fig. \ref{fig7}(d), which implies case $T_{6}$ points out the CP conservation.
 
From the analysis, out of six cases, only $T_{1}$ is found to be inconsistent with the experimental data for both normal and inverted mass ordering. For remaining cases the parameter space of CP violating phases ($\delta, \sigma, \rho$), effective mass term $|M|_{ee}$, neutrino masses ($m_{1}, m_{2}, m_{3}$) is found to be constrained to an appreciable extent at 3$\sigma$ CL.  The allowed ranges of all the five viable cases for  Dirac CP violating phase ($\delta$), Majorana phases ($\rho,\sigma$) and neutrino masses ($m_{1}, m_{2}, m_{3}$) have been summarized in Table \ref{tab3}.

\section{Summary and conclusion}

In the present work, we have systematically analyzed the texture one zero Majorana mass matrix along with zero trace condition.  In our analysis, we find that case $T_{1}$ with vanishing (1, 1) element of $M_{\nu}$ is ruled out with current experimental data. Therefore, out of six possible cases of texture one zero with zero trace, only five viz. $T_{2}, T_{3}, T_{4}, T_{5}, T_{6}$ can survive the current experimental tests at 3$\sigma$ CL. The ongoing and future neutrino based experiments including neutrinoless double beta decay  and cosmological experiments would test the validity of present texture zero analysis. 

\section*{Acknowledgment}

The author would like to thank the Director, National Institute of Technology Kurukshetra, for providing the necessary facilities to work.  \\

 \begin{table}
\begin{center}
\begin{footnotesize}
\resizebox{10cm}{!}{
\begin{tabular}{|c|c|}
  \hline
  
  Cases & Analytical expressions for $\xi$ and $\zeta$  \\

  \hline
  &\\
  $T_{1}$ & $\xi= +sec 2\theta_{12}(s_{12}^{2}- t_{13}^{2})$ \\
          & $\zeta =- sec 2\theta_{12}(c_{12}^{2}- t_{13}^{2})$\\
  &\\
  \hline
  &\\
  $T_{2}$ &  $\xi= \frac{(s_{12}^{2}s_{13}^{2}-c_{13}^{2})s_{23}^{2}+c_{12}c_{23}(c_{12}c_{23}e^{-i\delta}-2s_{12}s_{23}s_{13})e^{-i\delta}} {(s_{23}^{2}s_{13}^{2}-c_{23}^{2}e^{-2i\delta})c_{2(12)}+s_{2(12)}s_{2(23)}s_{13} e^{-i\delta}}$\\
   &\\
  &$\zeta= \frac{(-c_{12}^{2}s_{13}^{2}+c_{13}^{2})s_{23}^{2}-s_{12}c_{23}(s_{12}c_{23}e^{-i\delta}+2c_{12}s_{23}s_{13})e^{-i\delta}} {(s_{23}^{2}s_{13}^{2}-c_{23}^{2}e^{-2i\delta})c_{2(12)}+s_{2(12)}s_{2(23)}s_{13} e^{-i\delta}}$\\
  &\\
   \hline
   &\\
  $T_{3}$ &  $\xi=\frac{(s_{12}^{2}s_{13}^{2}-c_{13}^{2})c_{23}^{2}+c_{12}s_{23}(c_{12}s_{23}e^{-i\delta}+2s_{12}c_{23}s_{13})e^{-i\delta}} {(c_{23}^{2}s_{13}^{2}-s_{23}^{2}e^{-2i\delta})c_{2(12)}-s_{2(12)}s_{2(23)}s_{13} e^{-i\delta}}$ \\
  &$\zeta=\frac{(-c_{12}^{2}s_{13}^{2}+c_{13}^{2})c_{23}^{2}-s_{12}s_{23}(s_{12}s_{23}e^{-i\delta}-2c_{12}c_{23}s_{13})e^{-i\delta}} {(c_{23}^{2}s_{13}^{2}-s_{23}^{2}e^{-2i\delta})c_{2(12)}-s_{2(12)}s_{2(23)}s_{13} e^{-i\delta}}$\\
  &\\
   \hline
  
  &\\
  $T_{4}$ &  $\xi=\frac{-s_{12}c_{12}c_{23}e^{-i\delta}+s_{23}s_{13}(1+s_{12}^{2})}{s_{23}s_{13}(c_{12}^{2}-s_{12}^{2})+2s_{12}c_{12}c_{23}e^{-i\delta}}$\\
   &  $\zeta=\frac{-s_{12}c_{12}c_{23}e^{-i\delta}-s_{23}s_{13}(1+c_{12}^{2})}{s_{23}s_{13}(c_{12}^{2}-s_{12}^{2})+2s_{12}c_{12}c_{23}e^{-i\delta}}$\\
   &\\
  \hline
  &\\
  $T_{5}$ &  $\xi=\frac{s_{12}c_{12}s_{23}e^{-i\delta}+c_{23}s_{13}(1+s_{12}^{2})}{c_{23}s_{13}(c_{12}^{2}-s_{12}^{2})-2s_{12}c_{12}s_{23}e^{-i\delta}}$\\
   &  $\zeta=\frac{s_{12}c_{12}s_{23}e^{-i\delta}-c_{23}s_{13}(1+c_{12}^{2})}{c_{23}s_{13}(c_{12}^{2}-s_{12}^{2})-2s_{12}c_{12}s_{23}e^{-i\delta}}$\\
   &\\
  \hline
  &\\
   $T_{6}$ &  $\xi=\frac{s_{23}c_{23}(s_{12}^{2}s_{13}^{2}-c_{13}^{2}-c_{12}^{2}e^{-2i\delta})-c_{12}s_{12}s_{13}( c_{23}^{2}-s_{23}^{2})e^{-i\delta}}{s_{23}c_{23}(s_{13}^{2}+e^{-2i\delta})c_{2(12)}e^{-i\delta}+ 2s_{12}c_{12}s_{13}c_{2(23)}e^{-i\delta}}$ \\
   &  $\zeta=\frac{s_{23}c_{23}(-c_{12}^{2}s_{13}^{2}+c_{13}^{2}+s_{12}^{2}e^{-2i\delta})-c_{12}s_{12}s_{13}( c_{23}^{2}- s_{23}^{2})e^{-i\delta}}{s_{23}c_{23}(s_{13}^{2}+e^{-2i\delta})c_{2(12)}e^{-i\delta}+ 2s_{12}c_{12}s_{13}c_{2(23)}e^{-i\delta}}$ \\
  &\\  
 \hline
  \end{tabular}}
   \caption{ \label{tab2} The exact expressions of neutrino mass ratios $\xi $ and $\zeta$  of all the six one zero textures with vanishing trace is shown. The symbols $c_{2(ij)}$ $\equiv$ cos 2$\theta_{ij}$, $s_{2(ij)}$ $\equiv$ sin 2$\theta_{ij}$ are defined.}
   \end{footnotesize}
\end{center}

\end{table}
\newpage

\begin{table}[ht]
\begin{small}
\begin{center}
\begin{footnotesize}
\resizebox{14cm}{!}{
\begin{tabular}{|c|c|c|c|c|}
  \hline
Cases &\multicolumn{2}{c|}{Normal mass ordering
(NO)} &\multicolumn{2}{c|}{Inverted mass ordering
(IO)} \\ 
\hline $T_{1}$  & $\times$ & $\times$
&$\times$&$\times$  \\ & $\times$ &
$\times$ & $\times $ & $\times$ \\
&$\times$ & $\times$
& $\times$ & $\times$ \\ 
\hline 
$T_{2}$  &
$\delta= 153.7^{0}-158.4^{0} \oplus 204.1^{0}-206.3^{0} $& $m_{1}=0.0680-0.311$ &
$\delta= 0^{0}-21.06^{0} \oplus 126^{0}-156^{0} $ &$m_{1}=0.0425-0.357$ \\
& $\rho= -64.9^{0}--58.5^{0} \oplus 58.6^{0}-64.3^{0} $&$m_{2}=0.0712-0.314$ &$\oplus $&$m_{2}=0.0435-0.359$\\
&$\sigma= -64.5^{0}--58.89^{0} \oplus 57.8^{0}-64.3^{0} $ &$m_{3}=0.0842-0.343$&$204^{0}-233^{0}$&$m_{3}=0.00098-0.359$\\
&&&$\oplus $&\\
&&& $340^{0}-360^{0}$&\\
 && & $\rho= -60^{0}-60^{0} $ &  \\
&&
& $\sigma= -90^{0}--48.1^{0} \oplus 48.2^{0}-90^{0} $ &  \\ 
\hline 
$T_{3}$  &
$\delta= 12.49^{0}-25.9^{0}$ $\oplus$
$334.9^{0}-348.8^{0}$&  $m_{1}=0.0291-0.690$ &
$\delta= 23.7^{0}-50.48^{0}$  &$m_{1}=0.0422-0.377$ \\ 
&$\rho= -71.49^{0}--60.3^{0}$ $\oplus$
$59.36^{0}-69.7^{0}$ &$m_{2}=0.0307-0.690$&$\oplus$&$m_{2}=0.0435-0.377$ \\
&$\sigma=-73.34^{0}--59.62^{0}$ $\oplus$
$57.87^{0}-72.7^{0}$&$m_{3}=0.0480-0.690$&$159.6^{0}-201.2^{0}$&$m_{3}=0.00095-0.377$\\ 
 &&&$\oplus$&\\
&&&$309.2^{0}-338.1^{0}$&\\
&&&$\rho=-60^{0}-60^{0}$&  \\ 
&&& $\sigma= -90^{0}--48^{0}$ $\oplus$ $ 48^{0}-90^{0}$&\\ 
\hline

$T_{4}$  & $\delta= 86.68^{0}-94.95^{0}$ $\oplus$
$264.8^{0}-273.3^{0}$& $m_{1}=0.0269-0.0406$& $\times$ &
$\times$  \\ & $\rho=
-82.87^{0}--71.29^{0}$ $\oplus$ $69.57^{0}-82.08^{0}$
& $m_{2}=0.0280-0.0417$ & $\times$
&$\times$ \\ & $\sigma=
-82.87^{0}--71.7^{0}$ $\oplus$ $70.5^{0}-82.08^{0}$ &
$m_{3}=0.0516-0.0682$  & $\times$ & $\times$
\\ \hline

$T_{5}$  & $\delta= 85.99^{0}-94.48^{0}$ $\oplus$
$266.45^{0}-274.49^{0}$& $m_{1}=0.0269-0.0406$& $\times$ &
$\times$  \\ & $\rho=
-82.87^{0}--72.22^{0}$ $\oplus$ $71.88^{0}-83.01^{0}$
& $m_{2}=0.0280-0.0417$ & $\times$
&$\times$ \\ & $\sigma=
-82.87^{0}--72.22^{0}$ $\oplus$ $71.88^{0}-83.01^{0}$ &
$m_{3}=0.0516-0.0682$  & $\times$ & $\times$ \\ \hline

$T_{6}$  & $\times$&$\times$ &
$\delta= 53.25^{0}-68.5^{0}$  &$m_{1}=0.0449-0.0664$   \\ 
&$\times$&$\times$&$\oplus$&$m_{2}=0.0456-0.0672$\\
&$\times$&$\times$&$110.2^{0}-125^{0}$&$m_{3}=0.0099-0.0400$\\
&$\times$&$\times$& $\oplus$&\\
&$\times$&$\times$&$235.7^{0}-250^{0}$&\\
&$\times$&$\times$& $\oplus$&\\
&$\times$&$\times$&$291.3^{0}-309^{0}$&\\

&&& $\rho=
-54.73^{0}--44.7^{0}$ $\oplus$ $45.36^{0}- 54.5^{0}$ &\\ 
&&&$\sigma=-54.73^{0}--44.7^{0}$ $\oplus$ $45.36^{0}- 54.5^{0}$&  \\ 
\hline

\end{tabular}}

\caption{\label{tab3}The allowed ranges of
Dirac like CP violating phase $\delta$, the
Majorana phases $\rho, \sigma$ and three neutrino
masses $m_{1}, m_{2}, m_{3}$ for the
experimentally viable cases at 3$\sigma$ CL. Masses are in
eV.}
\end{footnotesize}
\end{center}
\end{small}
\end{table}

\end{document}